\documentclass[12pt,a4paper]{article}

\usepackage{amsmath}
\usepackage{amsfonts}
\usepackage{amssymb}
\usepackage{graphicx}
\usepackage{color}
\usepackage{booktabs}
\usepackage{inputenc}
\usepackage[T1]{fontenc}
\usepackage{mathrsfs}
\usepackage{enumerate}
\usepackage{xcolor,url}
\usepackage{cancel,dsfont} 

\newcommand{\eea}{\end{eqnarray}}
\newcommand{\bea}{\begin{eqnarray}}

\def\be{\begin{equation}}
\def\ee{\end{equation}}

\def\nn{\nonumber}

\newcommand{\de}{\partial}

\renewcommand{\(}{\left(}
\renewcommand{\)}{\right)}

\newcommand{\al}{\alpha}
\newcommand{\bt}{\beta}

\newcommand{\Om}{\Omega}

\def\hinvMpc{h\,{\rm Mpc}^{-1}}

\newcommand{\kmax}{k_{\rm max }}

\newcommand{\code}[1]{\texttt{#1}}

\usepackage[colorlinks,bookmarks]{hyperref}
\definecolor{linkblue}{rgb}{0,0,0.8}
\definecolor{linkgreen}{rgb}{0,0.5,0}

\hypersetup{pdfpagemode=UseNone, pdfstartview=FitH, linkcolor=linkblue,
            citecolor=linkgreen, urlcolor=linkblue}

\def\k{\mathbf{k}}
\def\q{\mathbf{q}}
\def\z{\mathbf{z}}

\begin{document}

\begin{center}

{\Large \bf {Limits on $w$CDM from the EFTofLSS  \\[0.3cm] 
with the PyBird code}
}
\\[0.7cm]

{\large  Guido D'Amico${}^{1}$,  Leonardo Senatore${}^{2,3}$,  Pierre Zhang${}^{4,5,6}$\\[0.7cm]}

\end{center}

\begin{center}

\vspace{.0cm}

{\normalsize { \sl $^{1}$ Dipartimento di SMFI dell' Universita' di Parma \& INFN Gruppo Collegato di Parma, Parma, Italy}}\\
\vspace{.3cm}

{\normalsize { \sl $^{2}$ Stanford Institute for Theoretical Physics, Physics Department,\\ Stanford University, Stanford, CA 94306}}\\
\vspace{.3cm}

{\normalsize { \sl $^{3}$ 
Kavli Institute for Particle Astrophysics and Cosmology,\\
 SLAC and Stanford University, Menlo Park, CA 94025}}\\
\vspace{.3cm}

{\normalsize { \sl $^{4}$ Department of Astronomy, School of Physical Sciences, \\
University of Science and Technology of China, Hefei, Anhui 230026, China}}\\
\vspace{.3cm}

{\normalsize { \sl $^{5}$ CAS Key Laboratory for Research in Galaxies and Cosmology, \\
University of Science and Technology of China, Hefei, Anhui 230026, China}}\\
\vspace{.3cm}

{\normalsize { \sl $^{6}$ School of Astronomy and Space Science, \\
University of Science and Technology of China, Hefei, Anhui 230026, China}}\\
\vspace{.3cm}

\vspace{.3cm}

\end{center}

\hrule \vspace{0.3cm}
{\small  \noindent \textbf{Abstract} We apply the Effective Field Theory of Large-Scale Structure to analyze the $w$CDM cosmological model.
By using the full shape of the power spectrum and the BAO post-reconstruction measurements from BOSS, the Supernovae from Pantheon, and a prior from BBN, we set the competitive CMB-independent limit $w=-1.046_{-0.052}^{+0.055}$ at $68\%$ C.L..
After adding the Planck CMB data, we find $w=-1.023_{-0.030}^{+0.033}$  at $68\%$ C.L..
Our results are obtained using PyBird, a new, fast Python-based code which we make publicly available.

\noindent

\vspace{0.3cm}}
\hrule

\vspace{0.3cm}
\newpage


\section{Introduction and Summary\label{sec:intro}}

\paragraph{Introduction} After a long journey, the Effective Field Theory of Large-Scale Structure (EFtofLSS) has been recently applied to the power spectrum of the galaxies of BOSS/SDSS~\cite{DAmico:2019fhj,Ivanov:2019pdj,Colas:2019ret}~\footnote{{Notice that Ref.~\cite{Colas:2019ret} is a companion paper to~\cite{DAmico:2019fhj}.} Ref.~\cite{DAmico:2019fhj} also applied it to the bispectrum, but finding marginal improvements, probably due to the fact that only the tree-level prediction was being used, so that the $k$-reach was not quite high.}. {These results have allowed us to measure all the cosmological parameters of the $\nu\Lambda$CDM model, except neutrino masses, just using a prior from Big Bang Nucleosynthesis (BBN).} The smallness of the error bars on some of these parameters have shown the power of Large-Scale Structure (LSS) surveys even without the inclusion of any cosmic microwave background (CMB) prior. For example, the constraint on the present-day dark matter fraction, $\Omega_m$, is competitive with the one from Planck2018~\cite{Aghanim:2018eyx}, and the one on the present-day Hubble parameter, $H_0$ is measured with the same precision as the one measured from Cosmic Distance Ladder, such as SH0ES~\cite{Riess:2019cxk}. Since the results from~\cite{DAmico:2019fhj,Ivanov:2019pdj,Colas:2019ret} are compatible with Planck, they contribute to shed light on the so-called Hubble tension (see a review in~\cite{Verde:2019ivm}). Though the number of modes in BOSS is much smaller than the ones in Planck, the origin of these remarkable results lies mainly in the fact that the CMB and LSS observables depend {quite} differently on the cosmological parameters, so that degeneracies are different (see for example sec.~4.3 of~\cite{DAmico:2019fhj}). Most importantly, these results show that the contribution of next generation LSS surveys, once analyzed with a controlled theory such as the EFTofLSS, to our  understanding of the history of the universe might be much larger than what previously believed, potentially helping to continue the remarkable exploration that was achieved in the past decades.

As mentioned, the application of the EFTofLSS to data is the result of a long journey where each of the ingredients of the EFTofLSS that was required in order to be able to apply it to data was one-by-one subsequently developed, tested on simulations, and shown to be successful. Though not all those intermediate results are directly used in the analysis, they were {\it necessary} for us, and probably for anybody else, to apply the model to data. We therefore find it fair, in each instance where the EFTofLSS is applied to data, to add the following footnote where we acknowledge at least a fraction of those most important developments~\footnote{The initial formulation of the EFTofLSS was performed in Eulerian space in~\cite{Baumann:2010tm,Carrasco:2012cv}, and subsequently extended to Lagrangian space in~\cite{Porto:2013qua}. The dark matter power spectrum has been computed at one-, two- and three-loop orders in~\cite{Carrasco:2012cv, Carrasco:2013sva, Carrasco:2013mua, Carroll:2013oxa, Senatore:2014via, Baldauf:2015zga, Foreman:2015lca, Baldauf:2015aha, Cataneo:2016suz, Lewandowski:2017kes,Konstandin:2019bay}. These calculations were accompanied by some  theoretical developments of the EFTofLSS, such as a careful understanding of renormalization~\cite{Carrasco:2012cv,Pajer:2013jj,Abolhasani:2015mra} (including rather-subtle aspects such as lattice-running~\cite{Carrasco:2012cv} and a better understanding of the velocity field~\cite{Carrasco:2013sva,Mercolli:2013bsa}), of the several ways for extracting the value of the counterterms from simulations~\cite{Carrasco:2012cv,McQuinn:2015tva}, and of the non-locality in time of the EFTofLSS~\cite{Carrasco:2013sva, Carroll:2013oxa,Senatore:2014eva}. These theoretical explorations also include an enlightening study in 1+1 dimensions~\cite{McQuinn:2015tva}. An IR-resummation of the long displacement fields had to be performed in order to reproduce the BAO peak, giving rise to the so-called IR-Resummed EFTofLSS~\cite{Senatore:2014vja,Baldauf:2015xfa,Senatore:2017pbn,Lewandowski:2018ywf,Blas:2016sfa}.  An account for baryonic effects was presented in~\cite{Lewandowski:2014rca}. The dark-matter bispectrum has been computed at one-loop in~\cite{Angulo:2014tfa, Baldauf:2014qfa}, the one-loop trispectrum in~\cite{Bertolini:2016bmt},
the displacement field in~\cite{Baldauf:2015tla}. The lensing power spectrum has been computed at two loops in~\cite{Foreman:2015uva}.  Biased tracers, such as halos and galaxies, have been studied in the context of the EFTofLSS in~\cite{ Senatore:2014eva, Mirbabayi:2014zca, Angulo:2015eqa, Fujita:2016dne, Perko:2016puo, Nadler:2017qto} (see also~\cite{McDonald:2009dh}), the halo and matter power spectra and bispectra (including all cross correlations) in~\cite{Senatore:2014eva, Angulo:2015eqa}. Redshift space distortions have been developed in~\cite{Senatore:2014vja, Lewandowski:2015ziq,Perko:2016puo}. Neutrinos has been included in the EFTofLSS in~\cite{Senatore:2017hyk,deBelsunce:2018xtd}, clustering dark energy in~\cite{Lewandowski:2016yce,Lewandowski:2017kes,Cusin:2017wjg,Bose:2018orj}, and primordial non-Gaussianities in~\cite{Angulo:2015eqa, Assassi:2015jqa, Assassi:2015fma, Bertolini:2015fya, Lewandowski:2015ziq, Bertolini:2016hxg}. Faster evaluation schemes for evaluation for some of the loop integrals have been developed in~\cite{Simonovic:2017mhp}.}.

\paragraph{Data sets} In this paper we focus on applying the EFTofLSS to analyze the $w$CDM model. We analyze various combinations among the full shape (FS) of BOSS DR12 pre-reconstructed power spectrum measurements  \cite{Gil-Marin:2015sqa}, baryon acoustic oscillations (BAO) of BOSS DR12 post-reconstructed power spectrum measurements \cite{Gil-Marin:2015nqa}, Planck2018 TT,TE,EE+lowE + lensing~\cite{Aghanim:2018eyx}. We also consider combinations with Supernovae (SN) measurements from the Pantheon Sample \cite{Scolnic:2017caz}. When quoting BAO, we also include measurements at small redshift from 6DF~\cite{Beutler:2011hx} and SDSS DR7 MGS \cite{Ross:2014qpa}, {as well as high redshift Lyman-$\alpha$ forest auto-correlation and cross-correlation with quasars from eBOSS DR14 measurements \cite{Agathe:2019vsu, Blomqvist:2019rah}.}
The inclusion of post-reconstructed BAO measurements gives a non-negligible improvement because the reconstruction amounts to using higher $n$-point functions. However the pre- and post-reconstruction BAO measurements are correlated. We describe how we account for this in App.~\ref{app:fsbao} {(see also~\cite{Philcox:2020vvt})}. When combined with Planck or SN, we simply add the {log-}likelihoods, since all the measurements refer to separate redshift bins. There is a small cross-correlation of the galaxy clustering data with the Planck weak lensing and the integrated Sachs-Wolfe (ISW) effect, which we neglect.

\paragraph{Methodology} The BOSS FS is analyzed following~\cite{DAmico:2019fhj,Colas:2019ret}. 
The description of the theory model, as well as the likelihood analysis, including the covariances and priors used, can be found there.
Let us give a quick summary of our methodology.
The theory model is the galaxy power spectrum in redshift space at one loop in the EFTofLSS~\cite{Perko:2016puo, DAmico:2019fhj}.
The power spectrum is properly IR-resummed~\cite{Senatore:2014vja,Senatore:2017pbn,Lewandowski:2018ywf}, and includes observational systematics corrections: the Alcock-Paszynski effect~\cite{Alcock:1979mp}, window functions~\cite{Beutler:2018vpe}, and fiber collisions~\cite{Hahn:2016kiy}.
The EFT parameters are given physical priors as discussed in~\cite{DAmico:2019fhj}. 	
We sample over all $w$CDM parameters: the baryon abundance, $\omega_b$, the cold dark matter abundance, $\omega_{\rm cdm}$, the Hubble constant, $H_0$, the amplitude of the primordial fluctuations, $\ln (10^{10} A_s)$, the tilt of the primordial power spectrum, $n_s$, and the dark energy equation of state parameter, $w$, imposing no prior on the cosmological parameters but a BBN prior on $\omega_b$ (see main text for more details).

App.~\ref{app:fsbao} provides checks on the joint analysis of the FS with BAO, and App.~\ref{app:theoryerror} estimates the theoretical error of the model in $w$CDM: while the theory-systematic errors has been measured for $\Lambda$CDM in~\cite{DAmico:2019fhj,Colas:2019ret} using large-volume N-body simulations, and App.~\ref{app:theoryerror} presents an analogous measurement for the EFTofLSS on $w$CDM.

We find that BOSS pre-reconstructed and post-reconstructed data can be analyzed up to, respectively, $\kmax = 0.23 \hinvMpc$ and $\kmax = 0.3 \hinvMpc$, using the same model, likelihood, priors on EFT parameters, and corrections to observational effects as described in~\cite{DAmico:2019fhj}.
This is because we measure the theory-systematic errors in all cosmological parameters by fitting the simulations on $w$CDM with a BBN prior, and we find them to be negligible once compared to the error bars obtained in the various analyses presented in this paper.

\begin{figure}
\centering
\includegraphics[width=0.54\textwidth]{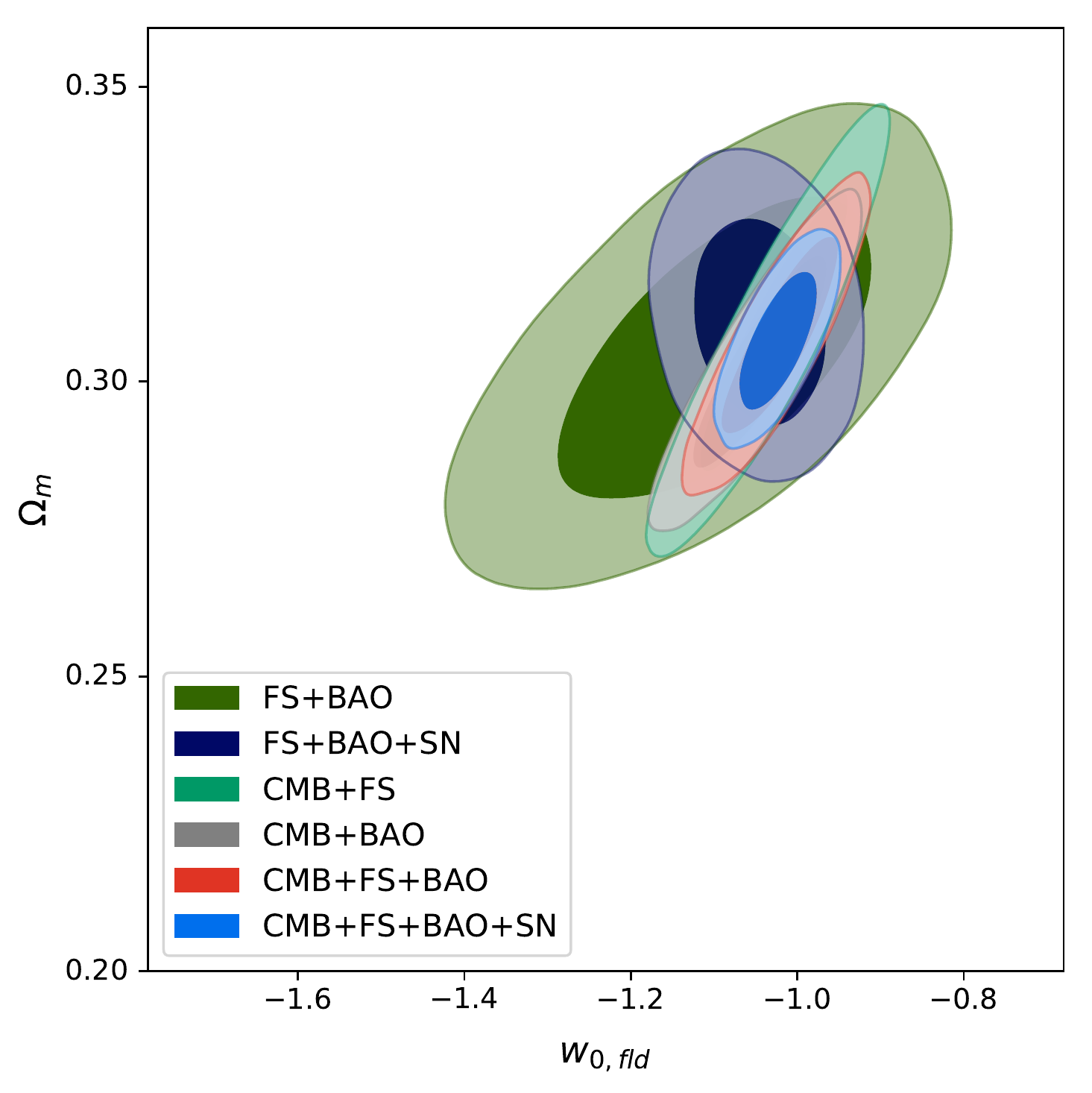}
\caption{ \label{fig:wcdm_w-omega_m} $w-\Omega_m$ contour from the various analyses performed in this work. {When not analyzed in combination with CMB, we always use a BBN prior.} These results show the power of LSS, when analyzed with the EFTofLSS approach at long wavelengths, to constrain dark energy.}
\end{figure}

\paragraph{Main Results} The main results of our analysis are maybe best represented by Fig.~\ref{fig:wcdm_w-omega_m}. {Using late-time measurements only, FS+BAO+SN, but with a BBN prior on the baryons abundance, we obtain a tight bound: $w=-1.046_{-0.052}^{+0.055}$ at $68\%$ C.L. This is looser than, but nevertheless competitive with, Planck2018~\cite{Aghanim:2018eyx} combined with other probes. Additionally, our limits appear to be an improvement with respect to DES results when not using CMB information~\cite{Abbott:2017wau,Abbott:2018wzc}, though it is difficult to perform a precise and quantitative comparison in this case due to different combinations of external data sets and priors being used.  It is hard to make a direct comparison with former BOSS analyses~\cite{Alam:2016hwk} as well, because in this case their results are presented always in combination with Planck. 
All products, plots, confidence intervals, results, etc., of our analyses are shown in sec.~\ref{sec:results}, and physical explanations on how all parameters from $w$CDM (but in fact several extensions of $\Lambda$CDM) can be measured independently using FS+BAO are discussed in sec.~\ref{sec:physics}.  Fitting the combination of all datasets, we obtain: $w=-1.023_{-0.030}^{+0.033}$.}

In our analysis, we have assumed that the BOSS data do not contain any residual systematic error. Given the additional cosmological information that the EFTofLSS enables us to exploit from these data, it might be important to investigate if potential undetected systematic errors can affect our results. We leave such investigation to future work.

\paragraph{Public Fast Code} Details on the code used to fit the FS, called PyBird: Python code for Biased tracers in redshift space, are given in sec.~\ref{sec:pybird}. In fact, an additional result of this paper is making available to the general community a fast and simple python-based code to perform the FS analysis of the power spectrum using the EFTofLSS.  Pybird can be found at \href{https://github.com/pierrexyz/pybird}{https://github.com/pierrexyz/pybird}, {where we also provide an explicit likelihood to be used in the MonteCarlo sampler MontePython~\cite{Brinckmann:2018cvx, Audren:2012wb}}.

\section{Results on $w$CDM}
\label{sec:results}

\subsection{FS + BAO (+ SN)}
Here we present our results on the CMB-independent analysis of the $w$CDM model with fixed neutrino masses.
The datasets we use are the following:
\begin{itemize}
    \item{FS} refers to the combination of the power spectra (monopole and quadrupole) of the three different sky-cuts CMASS NGC, CMASS SGC and LOWZ NGC.
    These are at the effective redshift $z_{\rm eff} = 0.57$ for CMASS and $z_{\rm eff} = 0.32$ for LOWZ, and the maximum wavenumber we consider is $k_{\rm max} = 0.23 h/{\rm Mpc}$ for CMASS and $k_{\rm max} = 0.20 h/{\rm Mpc}$ for LOWZ.  Ref.~\cite{DAmico:2019fhj,Colas:2019ret} showed that in $\Lambda$CDM, with this choice of $k_{\rm max}$, the theoretical systematic error is negligible for this data set with the theoretical model that we use. More explanations on this analysis are given at the beginning of appendix~\ref{app:fsbao}.
    
    \item{FS+BAO} refers to the combination of the previous dataset with the $H r_s$ and $D_A / r_s$ parameters measured from the post-reconstructed power spectra corresponding to the same sky-cuts.
    We include the covariance among these datasets calculated as explained in appendix~\ref{app:fsbao}.
    In addition, we add `small-$z$' BAO measurements at redshift $z_{\rm eff} = 0.106$ from 6DF and $z_{\rm eff} = 0.15$ from SDSS DR7 MGS, {as well as Ly$\alpha$ BAO measurements including auto-correlation and cross-correlation with quasars from eBOSS at $z_{\rm eff} = 2.34$ and $z_{\rm eff} = 2.35$ respectively, based on the likelihood of \cite{Cuceu:2019for}.}
    All these redshift bins are uncorrelated with the redshift bins of FS+BAO.
    \item{FS+BAO+SN} refers to the combination of the previous dataset plus the Pantheon catalogue of high-redshift supernovae.
\end{itemize}

We use a Gaussian prior on $\omega_b$ motivated from BBN constraints centered on $0.02235$ with $\sigma_{\rm BBN} = 0.0005$, which is obtained by adding up the theory and statistic error of~\cite{Cooke:2017cwo}. The fit is done considering the Planck prescription of one single massive neutrino with mass $0.06$ eV as done in \cite{Aghanim:2018eyx}. 
We assume Gaussian initial conditions, pure CDM with no WDM/HDM admixture, and a dark energy fluid component described by constant-in-time equation of state $p=w \rho$ and without perturbations.
This is the standard phenomenological dark energy model traditionally analyzed using LSS data and named $w$CDM (see e.g.~\cite{Alam:2016hwk,Abbott:2017wau,Abbott:2018wzc}).

\begin{figure}[h!]
\centering
\includegraphics[width=0.99\textwidth]{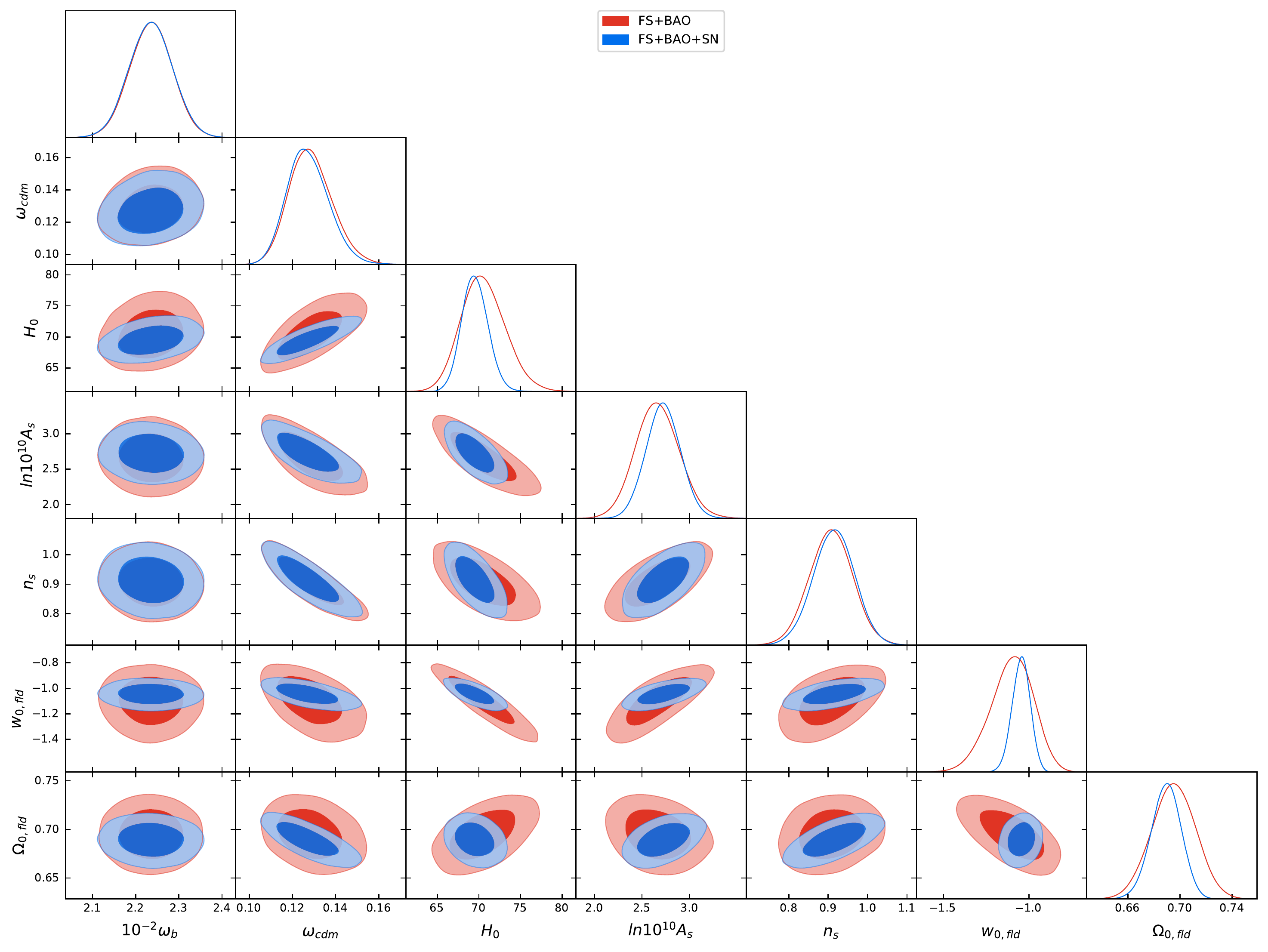}
\caption{ \label{fig:wcdm_lss} Triangle plot of $w$CDM fitting various combinations of FS with BAO and SN.}
\end{figure}

The best fits, means and one-sigma intervals of the 1D posteriors are given in Table~\ref{tab:wcdm_lss}. 
The triangle plots are shown in Fig.~\ref{fig:wcdm_lss}, while the $w-\Omega_m$ contour is shown in Fig.~\ref{fig:wcdm_w-omega_m}. {Following~\cite{DAmico:2019fhj}, we compare our analysis pipeline with simulations in App.~\ref{app:fsbao}, finding negligible theoretical systematic errors in $w$CDM.}

\begin{table}
\centering
\scriptsize
\begin{tabular}{|p{3cm}|p{2cm}|p{2cm}|} 
 \hline 
FS+BAO & best-fit & mean$\pm\sigma$  \\ \hline 
$w_{0,fld }$ &$-1.085$ & $-1.101_{-0.11}^{+0.14}$ \\ 
$100~\omega_{b }$ &$2.229$ & $2.236_{-0.05}^{+0.051}$  \\ 
$\omega_{cdm }$ &$0.122$ & $0.1286_{-0.011}^{+0.009}$ \\ 
$H_0$ &$69.68$ & $70.53_{-2.9}^{+2.4}$  \\ 
$\ln\left(10^{10}A_{s }\right)$ &$2.791$ & $2.664_{-0.24}^{+0.22}$  \\ 
$n_{s }$ &$0.9447$ & $0.9071_{-0.056}^{+0.057}$  \\ \hline
$\Omega_{0,fld }$ &$0.7014$ & $0.695_{-0.017}^{+0.018}$  \\ 
$\sigma_8$ &$0.7368$ & $0.7062_{-0.057}^{+0.049}$ \\ 
\hline 
 \end{tabular}
 \begin{tabular}{|p{3cm}|p{2cm}|p{2cm}|} 
 \hline 
FS+BAO+SN & best-fit & mean$\pm\sigma$ \\ \hline 
$w_{0,fld }$ &$-1.031$ & $-1.046_{-0.052}^{+0.055}$  \\  
$100~\omega_{b }$ &$2.235$ & $2.235_{-0.052}^{+0.051}$ \\ 
$\omega_{cdm }$ &$0.1218$ & $0.1273_{-0.011}^{+0.0088}$  \\ 
$H_0$ &$68.69$ & $69.52_{-1.7}^{+1.5}$ \\ 
$\ln\left(10^{10}A_{s }\right)$ &$2.777$ & $2.724_{-0.18}^{+0.19}$  \\ 
$n_{s }$ &$0.9273$ & $0.9144_{-0.053}^{+0.055}$  \\ \hline
$\Omega_{0,fld }$ &$0.6931$ & $0.6892_{-0.011}^{+0.012}$\\ 
$\sigma_8$ &$0.7144$ & $0.7155_{-0.051}^{+0.045}$  \\ 
\hline 
 \end{tabular} \\ 
\begin{tabular}{|p{3cm}|p{2cm}|p{2cm}|} 
 \hline 
FS+BAO {\tiny w/o Ly-$\alpha$}& best-fit & mean$\pm\sigma$ \\ \hline 
$w_{0,fld }$ &$-1.021$ & $-1.109_{-0.15}^{+0.17}$  \\ 
$100~\omega_{b }$ &$2.236$ & $2.239_{-0.051}^{+0.051}$  \\ 
$\omega_{cdm }$ &$0.1298$ & $0.1439_{-0.019}^{+0.013}$  \\ 
$H_0$ &$69.2$ & $71.88_{-3.7}^{+2.9}$  \\ 
$\ln\left(10^{10}A_{s }\right)$ &$2.701$ & $2.537_{-0.27}^{+0.24}$  \\ 
$n_{s }$ &$0.9045$ & $0.8433_{-0.067}^{+0.082}$ \\ \hline
$\Omega_{0,fld }$ &$0.6809$ & $0.6769_{-0.021}^{+0.025}$ \\ 
$\sigma_8$ &$0.7087$ & $0.6918_{-0.057}^{+0.049}$ \\ 
\hline 
\end{tabular}
\begin{tabular}{|p{3cm}|p{2cm}|p{2cm}|} 
\hline 
FS+BAO+SN {\tiny w/o Ly-$\alpha$}& best-fit & mean$\pm\sigma$  \\ \hline 
$w_{0,fld }$ &$-1.077$ & $-1.083_{-0.056}^{+0.07}$ \\ 
$100~\omega_{b }$ &$2.247$ & $2.238_{-0.05}^{+0.052}$  \\ 
$\omega_{cdm }$ &$0.1405$ & $0.1434_{-0.019}^{+0.012}$  \\ 
$H_0$ &$71.28$ & $71.44_{-2.5}^{+1.8}$  \\ 
$\ln\left(10^{10}A_{s }\right)$ &$2.656$ & $2.553_{-0.21}^{+0.2}$  \\ 
$n_{s }$ &$0.856$ & $0.845_{-0.061}^{+0.081}$  \\  \hline
$\Omega_{0,fld }$ &$0.6778$ & $0.6746_{-0.014}^{+0.017}$  \\ 
$\sigma_8$ &$0.7264$ & $0.6937_{-0.051}^{+0.043}$  \\ 
\hline 
\end{tabular}\\
\caption{\label{tab:wcdm_lss} Results on $w$CDM fitting various combinations of FS with BAO and SN.}
\end{table}

For our constraints on $w$CDM, we only show the results for FS + BAO and FS + BAO + SN, since the FS dataset alone does not constrain $w$ {very well, as this} is subject to strong degeneracies. This is to be expected, as we analyze only two close redshift bins with $z_{\rm eff} = 0.32$ and $z_{\rm eff} = 0.57$.
When adding BAO, $w$ is constrained to { $w=-1.101 \pm 0.12$ at $68\%$ C.L. ($\pm 0.25$ at $95\%$ C.L.).}
Much of the improvement is coming from the addition of the small-$z$ bins with $z_{\rm eff} = 0.106$ from 6DF and $z_{\rm eff} = 0.15$ from SDSS DR7 MGS at which dark energy has strong effect on the background evolution, allowing to {further} break the degeneracies. {Adding Ly-$\alpha$ BAO breaks {even more the} degeneracies, as it adds one more redshift point deep into matter domination. These combinations of FS plus the various BAO are shown in Fig.~\ref{fig:wcdm_fs-allbao}, and more discussions can be found in sec.~\ref{sec:physics}.}

Although {the result from FS+BAO} does not set a very strong constraint on $w$, it is interesting in itself since it shows the ability to measure dark energy evolution using only late-time observables.
Notice that the limits on $w$ from the BOSS (+SDSS/6DF) data are stronger than the ones from Planck2018 (+ lensing) alone, which measures $w$ to $-1.57_{-0.33}^{+0.16}$ at $68\%$ C.L. ($_{-0.40}^{+0.50}$ at $95\%$ C.L.)~\cite{Aghanim:2018eyx}.

The ability to measure $w$ with FS+BAO is to be contrasted with the SN measurements, which require an external input (usually from the CMB) to achieve competitive precision, to break the degeneracy line in the plane $w-\Omega_m$.
The SN constraint on $w-\Omega_m$ comes from the fact that they provide a measurement of the luminosity distance $D_L(z)$.
We can estimate the scaling on parameters at the median redshift of the SN sample, getting $D_L(z=0.25) \sim \Omega_m^{-0.055} |w|^{0.1}$, as can be inferred from table.~\ref{tab:derivatives}. 
This is to be contrasted with the positive correlation between $w$ and $\Omega_m$ given by the FS+BAO measurements, as discussed in sec.~\ref{sec:physics}.
Therefore, the combination FS+BAO+SN allows a much tighter constraint on {$w=-1.046_{-0.052}^{+0.055}$}.
This sets a competitive limit from late-time measurements alone, and it is consistent with previous analyses~\footnote{Notice that the change of the results after adding SN is statistically consistent, as shown in Table~\ref{tab:wcdm_lss} which includes the best fit values. Consistently, Fig.~\ref{fig:wcdm_w-omega_m} shows that there appears to be no statistical incompatibility between the data sets, also when, as we will do next, we add the CMB data.}.

For instance, combined Planck2018 and BAO gives $w = -1.038_{-0.048}^{+0.055}$~\cite{Aghanim:2018eyx}, while combined Planck2015 and SN yields $w = -1.026 \pm 0.041$~\cite{Scolnic:2017caz}.
This strongly suggests that additional data from future spectroscopic and photometric surveys will allow to constrain dark energy in an unprecedented way.

\begin{table}
\centering
\scriptsize
\begin{tabular}{|p{3cm}|p{2cm}|p{2cm}|} 
\hline 
CMB+BAO & best-fit & mean$\pm\sigma$ \\ \hline 
$w_{0,fld }$ &$-1.035$ & $-1.045_{-0.051}^{+0.056}$ \\ 
$100~\omega_{b }$ &$2.24$ & $2.238_{-0.014}^{+0.014}$  \\ 
$\omega_{cdm }$ &$0.12$ & $0.12_{-0.0011}^{+0.0011}$ \\ 
$100*\theta_{s }$ &$1.042$ & $1.042_{-0.00029}^{+0.0003}$  \\ 
$\ln\left(10^{10}A_{s }\right)$ &$3.051$ & $3.047_{-0.015}^{+0.014}$ \\ 
$n_{s }$ &$0.9667$ & $0.9658_{-0.0041}^{+0.0039}$ \\ 
$\tau_{reio }$ &$0.05767$ & $0.05514_{-0.0078}^{+0.0073}$ \\  \hline
$z_{reio }$ &$8.021$ & $7.751_{-0.76}^{+0.76}$ \\ 
$\Omega_{0,fld }$ &$0.6951$ & $0.697_{-0.012}^{+0.012}$ \\ 
$YHe$ &$0.2479$ & $0.2478_{-6.1e-05}^{+6.1e-05}$ \\ 
$H_0$ &$68.49$ & $68.75_{-1.5}^{+1.3}$ \\ 
$\sigma_8$ &$0.8239$ & $0.8244_{-0.017}^{+0.016}$ \\ 
\hline 
 \end{tabular}
 \begin{tabular}{|p{3cm}|p{2cm}|p{2cm}|} 
 \hline 
CMB+FS & best-fit & mean$\pm\sigma$ \\ \hline 
$w_{0,fld }$ &$-1.027$ & $-1.029_{-0.056}^{+0.063}$  \\ 
$100~\omega_{b }$ &$2.241$ & $2.238_{-0.014}^{+0.014}$  \\ 
$\omega_{cdm }$ &$0.1199$ & $0.12_{-0.0011}^{+0.0011}$ \\ 
$100*\theta_{s }$ &$1.042$ & $1.042_{-0.00028}^{+0.00029}$  \\ 
$\ln\left(10^{10}A_{s }\right)$ &$3.047$ & $3.044_{-0.015}^{+0.014}$  \\ 
$n_{s }$ &$0.9648$ & $0.9656_{-0.004}^{+0.004}$  \\ 
$\tau_{reio }$ &$0.05567$ & $0.05385_{-0.0077}^{+0.0073}$ \\  \hline
$z_{reio }$ &$7.817$ & $7.623_{-0.75}^{+0.77}$  \\ 
$\Omega_{0,fld }$ &$0.6935$ & $0.6923_{-0.016}^{+0.016}$ \\ 
$YHe$ &$0.2479$ & $0.2478_{-6.1e-05}^{+6.1e-05}$  \\ 
$H_0$ &$68.3$ & $68.26_{-1.8}^{+1.6}$  \\ 
$\sigma_8$ &$0.819$ & $0.8189_{-0.018}^{+0.016}$ \\ 
\hline 
 \end{tabular} \\
 \begin{tabular}{|p{3cm}|p{2cm}|p{2cm}|} 
 \hline 
CMB+FS+BAO & best-fit & mean$\pm\sigma$  \\ \hline 
$w_{0,fld }$ &$-1.018$ & $-1.021_{-0.044}^{+0.049}$  \\ 
$100~\omega_{b }$ &$2.238$ & $2.24_{-0.014}^{+0.014}$ \\ 
$\omega_{cdm }$ &$0.1198$ & $0.1197_{-0.0011}^{+0.0011}$  \\ 
$100*\theta_{s }$ &$1.042$ & $1.042_{-0.00029}^{+0.00029}$ \\ 
$\ln\left(10^{10}A_{s }\right)$ &$3.038$ & $3.045_{-0.015}^{+0.014}$  \\ 
$n_{s }$ &$0.9654$ & $0.9664_{-0.0039}^{+0.004}$ \\ 
$\tau_{reio }$ &$0.0526$ & $0.05492_{-0.0077}^{+0.0072}$  \\   \hline
$z_{reio }$ &$7.509$ & $7.722_{-0.74}^{+0.76}$ \\ 
$\Omega_{0,fld }$ &$0.6909$ & $0.6926_{-0.011}^{+0.012}$  \\ 
$YHe$ &$0.2478$ & $0.2479_{-6e-05}^{+5.9e-05}$ \\ 
$H_0$ &$67.98$ & $68.18_{-1.3}^{+1.2}$ \\ 
$\sigma_8$ &$0.8126$ & $0.8162_{-0.015}^{+0.014}$  \\ 
\hline 
 \end{tabular} 
 \begin{tabular}{|p{3cm}|p{2cm}|p{2cm}|} 
 \hline 
{ CMB+FS+BAO+SN} & best-fit & mean$\pm\sigma$  \\ \hline 
$w_{0,fld }$ &$-1.028$ & $-1.023_{-0.03}^{+0.033}$ \\ 
$100~\omega_{b }$ &$2.236$ & $2.24_{-0.014}^{+0.014}$  \\ 
$\omega_{cdm }$ &$0.1202$ & $0.1197_{-0.001}^{+0.0011}$ \\ 
$100*\theta_{s }$ &$1.042$ & $1.042_{-0.00029}^{+0.00029}$  \\ 
$\ln\left(10^{10}A_{s }\right)$ &$3.036$ & $3.045_{-0.015}^{+0.014}$ \\ 
$n_{s }$ &$0.9658$ & $0.9663_{-0.004}^{+0.0039}$  \\ 
$\tau_{reio }$ &$0.0507$ & $0.05464_{-0.0076}^{+0.0072}$  \\  \hline
$z_{reio }$ &$7.321$ & $7.694_{-0.73}^{+0.76}$ \\ 
$\Omega_{0,fld }$ &$0.6914$ & $0.6931_{-0.0078}^{+0.008}$  \\ 
$YHe$ &$0.2478$ & $0.2479_{-6e-05}^{+6e-05}$  \\ 
$H_0$ &$68.13$ & $68.22_{-0.86}^{+0.82}$  \\ 
$\sigma_8$ &$0.8161$ & $0.8168_{-0.011}^{+0.011}$ \\ 
\hline 
 \end{tabular} \\ 
 
\caption{\label{tab:wcdm_cmb} Results on $w$CDM fitting different combinations of CBM with FS, BAO and SN.}
\end{table}

\begin{figure}[h]
\centering
\includegraphics[width=0.99\textwidth]{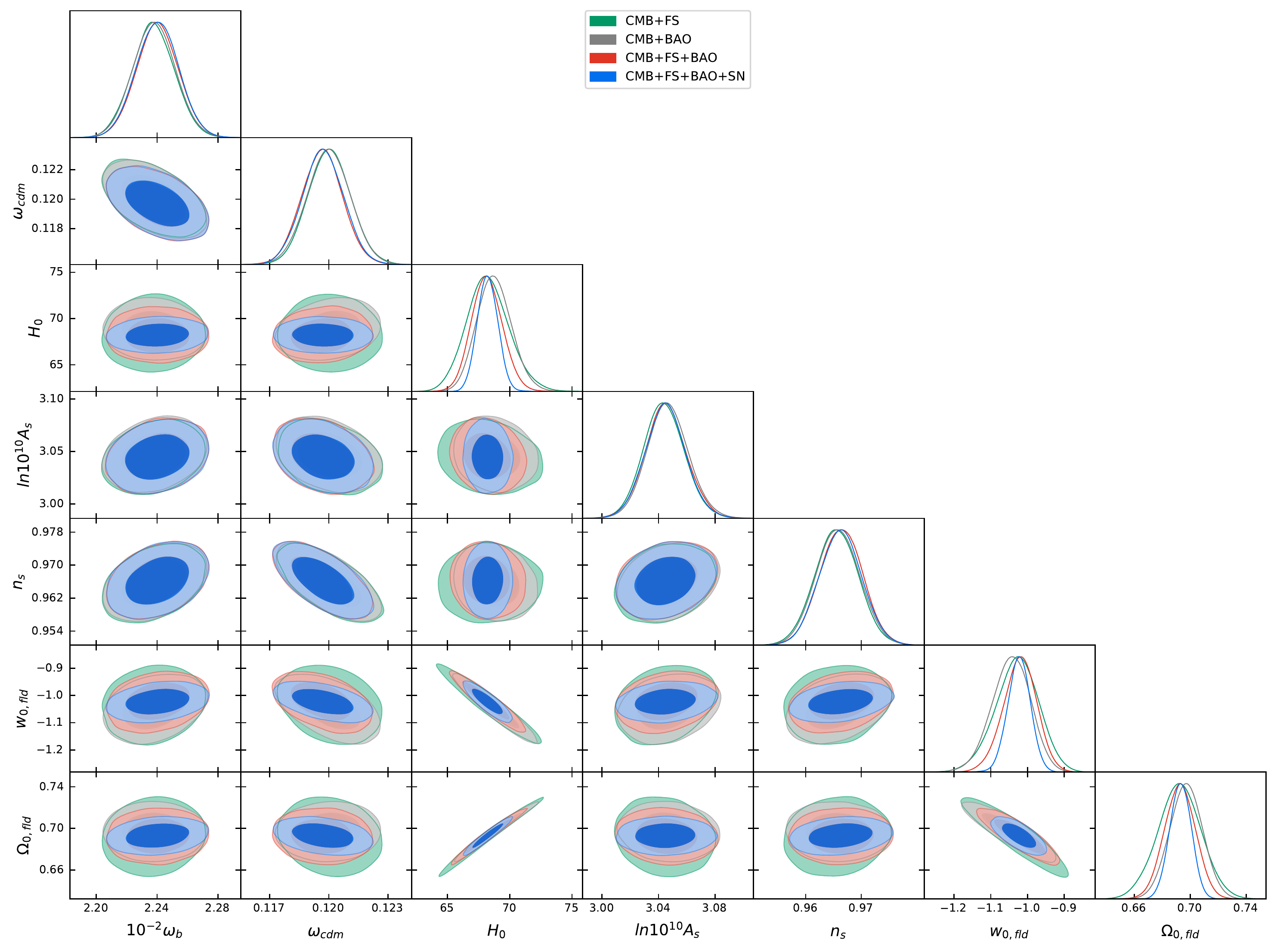}
\caption{ \label{fig:wcdm_cmb} Triangle plot of $w$CDM fitting various combinations of CMB with FS, BAO and SN.}
\end{figure}

\subsection{Combined CMB and FS + BAO (+ SN)}
In order to get the tightest constraints possible on $w-\Omega_m$, we add the Planck2018 datasets to the FS, BAO and SN.
When adding Planck2018, we include the temperature, polarization and lensing likelihoods as provided by the Planck collaboration, {that we will refer as `CMB'}. As for the nuisance parameters, we only consider the `lite' configuration with one nuisance parameter, since we verified that it gives very similar results with respect to the full configuration.
{When combined to CMB, we do not include Lyman-$\alpha$ BAO to facilitate comparison with other analyses. We checked that including them does not change the results.}
The best-fits, means and one-sigma intervals of the 1D posteriors are given in Table~\ref{tab:wcdm_cmb}.
The triangle plots are shown in Fig.~\ref{fig:wcdm_cmb}, and the $w-\Omega_m$ contour is shown in Fig.~\ref{fig:wcdm_w-omega_m}.

The results we get on CMB + BAO are similar to the one of~\cite{Aghanim:2018eyx}.
Our CMB + BAO constraint is $w=-1.045_{-0.051}^{+0.056}$, while CMB + FS gives $w=-1.029_{-0.056}^{+0.063}$, a similar constraint with a slight shift towards $-1$.
As argued in \cite{Ivanov:2019hqk} for the $\Lambda$CDM model, the similar error bars for the CMB + BAO and CMB + FS analyses are a coincidence given the BOSS volume and the reconstruction algorithm used to measure the BAO parameters: it is expected that the FS information will supersede the BAO information in the next-generation experiments.
At this stage, we note that both the FS and the BAO information break the degeneracy in the $w-\Omega_m$ plane displayed by the Planck fit alone.
The combination of CMB + FS + BAO gives an even tighter constraint: $w=-1.021_{-0.044}^{+0.049}$, which is about a $15\%$ improvement on the error bar compared to CMB + BAO.
This shows that the FS does add information on top of the BAO, even when in combination with CMB.
Finally, adding SN provides $w=-1.021_{-0.030}^{+0.033}$, our tightest constraint.
This combination gives similar error bars as CMB+BAO+SN measurements, $w=-1.028 \pm 0.031$~\cite{Aghanim:2018eyx}, but a slight shift towards $-1$.

\section{Physical Considerations}
\label{sec:physics}

Let us now try to understand analytically how the data allow us to break the degeneracies and give constraints on $w$, following the discussion in~\cite{DAmico:2019fhj}, which, in turns, is similar to what done for the CMB in~\cite{Percival:2002gq}.

\begin{figure}[h]
\centering
\includegraphics[width=0.99\textwidth]{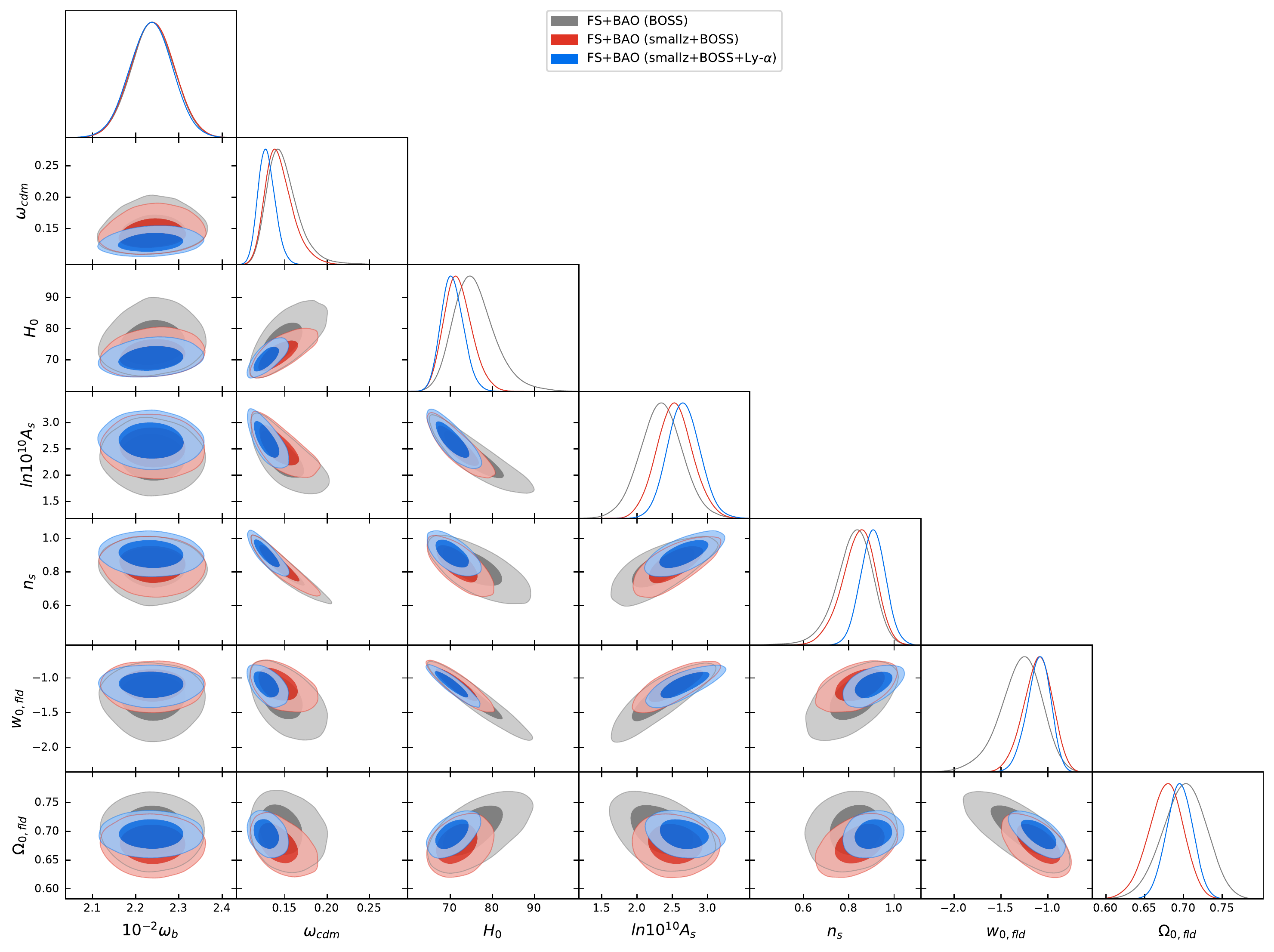}
\caption{Triangle plot of $w$CDM fitting the FS in combination with various BAO: BOSS, BOSS+small-$z$ (SDSS DR7 MGS/6DF) and BOSS+small-$z$+Lyman-$\alpha$.}
\label{fig:wcdm_fs-allbao}
\end{figure}

First, we notice that changing $w$ does not impact the shape of the primordial power spectrum.
The only effects are the modifications of the linear growth function and of the angular diameter distance, which affects the angular scale under which the BAO peak is seen.

\begin{table}
    \scriptsize
    \centering
    \begin{tabular}{|l|c|c|c|c|c|c|c|c|} 
    \hline 
    & \multicolumn{3}{c|}{$\theta_{\rm LSS, \, \perp}$}
    & \multicolumn{3}{c|}{$\theta_{\rm LSS, \, \parallel}$}
    & \multicolumn{2}{c|}{$\theta_{\rm LSS, \, V}$} \\
    \hline 
    &$z_{\rm CMASS}$ & $z_{\rm LOWZ}$ & $z_{\rm Ly-\alpha}$
    &$z_{\rm CMASS}$ & $z_{\rm LOWZ}$ & $z_{\rm Ly-\alpha}$
    &$z_{\rm 6dF}$ & $z_{\rm MGS}$ \\
    \hline
    $\omega_{m}$
    &$-0.14$ & $-0.19$ & $0.15$
    & $-0.027$ & $-0.12$ & $0.20$
    & $-0.23$ & $-0.21$ \\
    $h$
    &$0.77$ & $0.86$ & $0.45$
    & $0.54$ & $0.72$ & $0.84$
    & $0.94$ & $0.91$ \\
    $|w|$
    &$-0.17$ & $-0.12$ & $-0.19$
    & $-0.25$ & $-0.21$ & $-0.11$
    & $-0.065$ & $-0.087$ \\ 
    \hline 
    \end{tabular} 
    \begin{tabular}{|l|c|c|c|c|c|} 
        \hline 
        & \multicolumn{2}{c|}{$D$}
        & \multicolumn{2}{c|}{$f$}
        & $D_L$ \\
        \hline 
        &$z_{\rm CMASS}$ & $z_{\rm LOWZ}$
        &$z_{\rm CMASS}$ & $z_{\rm LOWZ}$
        &$z=0.25$ \\
        \hline
        $\omega_{m}$
        & $-0.12$ & $-0.08$
        & $0.30$ & $0.40$ 
        & $-0.055$ \\
        $h$
        & $0.25$ & $0.16$
        & $-0.59$ & $-0.80$
        & $ -0.89$ \\
        $|w|$
        & $-0.063$ & $-0.028$
        & $0.29$ & $0.25$
        & $0.10$ \\ 
        \hline 
        \end{tabular} 
\caption{Logarithmic derivatives of different observables with respect to cosmological parameters.}
\label{tab:derivatives}
\end{table}

BAOs mainly measure the combination $\theta_{\rm LSS, \, V} = \left( \theta_{\rm LSS, \, \perp}^2 \theta_{\rm LSS, \, \parallel}\right)^\frac{1}{3}$, where $\theta_{\rm LSS, \, \perp}$ and $\theta_{\rm LSS, \, \parallel}$ are the contributions perpendicular and parallel to the line of sight, which can be disentangled from measuring separately the monopole and the quadrupole of the power spectrum {in the FS analysis, or from measuring the anisotropic BAO parameters}.
The angles are the following ratios of length scales: 
\begin{equation}
    \theta_{\rm LSS, \, \perp} \simeq \frac{ r_d(z_{\rm CMB})}{D_A(z_{\rm LSS})} \, , \quad
    \theta_{\rm LSS,\, \parallel } \simeq \frac{ r_d(z_{\rm CMB})}{ c\, z_{\rm LSS}/H(z_{\rm LSS})} \, ,
\end{equation}
where $r_d(z_{\rm CMB})$ is the sound horizon at drag redshift, and $z_{LSS}$ is the mean redshift of the data sample.
Following~\cite{DAmico:2019fhj}, to understand the approximate parameter scaling, we  take the log derivatives around a fiducial cosmology ($\omega_m=0.147$, $h=0.7$, $w=-1$), shown in tab.~\ref{tab:derivatives} for several quantities of interest.

From the table, we can read the approximate degeneracy lines for CMASS, LOWZ, small-$z$ and Ly$\alpha$:
\bea\nn
&&\theta_{\rm LSS, \, \perp}(z_{\rm CMASS}) \sim \(h \, \omega_m^{-0.18} |w|^{-0.22} \)^{0.77},\qquad \theta_{\rm LSS, \, \parallel}(z_{\rm CMASS}) \sim \(h \, \omega_m^{-0.05} |w|^{-0.47}\)^{0.54},\\  \nn
&&\theta_{\rm LSS, \, \perp}(z_{\rm LOWZ}) \sim \(h \, \omega_m^{-0.22} |w|^{-0.14}\)^{0.86}, \qquad\theta_{\rm LSS, \, \parallel}(z_{\rm LOWZ}) \sim \(h \, \omega_m^{-0.16} |w|^{-0.29}\)^{0.72}\ ,\\ \nn
&&\theta_{\rm LSS, \, V}(z_{\rm 6dF}) \sim \( h \, \omega_m^{-0.25} |w|^{-0.069}\)^{0.94},\qquad \theta_{\rm LSS, \, V}(z_{\rm MGS}) \sim \(h \, \omega_m^{-0.23} |w|^{-0.096}\)^{0.91}\ , \\
&&\theta_{\rm LSS, \, \parallel}(z_{{\rm Ly}\alpha}) \sim \(h \, \omega_m^{2.4} |w|^{-1.3}\)^{0.84},\qquad \theta_{\rm LSS, \, \perp}(z_{{\rm Ly}\alpha}) \sim \(h \, \omega_m^{0.33} |w|^{-0.42}\)^{0.45}\ .
\eea

The relative amplitude of the BAO wiggles with respect to the smooth part instead gives a measurement of $\sim \omega_m$.
This is quite intuitive as, unlike the wavelength, the amplitude of the oscillating part is not affected by projection effects, and therefore simply scales as the density of baryons and dark matter at the time of recombination, which, in the case of fixed $\omega_b$, simply scales as $\omega_m$ (see~\cite{Mukhanov:2003xr} for a very pedagogical review and derivation).

Let us now consider the broadband signal.
The linear power spectrum monopole and quadrupole give a measurement of $b_1^2 A_s^{(\kmax)}$ and $b_1 f A_s^{(\kmax)}$, where $f(z_{\rm CMASS}) \sim \Om_m^{0.3} |w|^{0.29}$ and $f(z_{\rm LOWZ}) \sim \Om_m^{0.4} |w|^{0.25}$, $f$ being the log-derivative of the linear growth function.
Here $A_s^{(\kmax)}$ represents the amplitude of the linearly evolved power spectrum at the maximum wavenumber of our analysis, as this is where the signal peaks.
The broadband signal gives also a measurement of the linear growth function $D$ since the linear power spectrum scales as $D^2$, while the loop one scales as $D^4$.
However, the dependence on $w$ is quite weak, since $D(z_{\rm CMASS}) \sim \Omega_m^{-0.12} |w|^{-0.063}$ and $D(z_{\rm LOWZ}) \sim \Omega_m^{-0.08} |w|^{-0.028}$.

These estimates allow us, in principle, to solve for the cosmological parameters of the $w$CDM model when using only LSS data.
We determine $\omega_m$ from the amplitude of the BAO oscillations.
Then, the BAO angles, the quadrupole/monopole ratio and the amplitude of the broadband signal will give a measurement of $h$, $A_s$ and $w$ {(as well as of $b_1$)}.
However, using only the FS dataset there will be large error bars in the recovered parameters, given the low precision of the quadrupole spectra, and of the LOWZ measurements.
In particular, we find a large anticorrelation among $w$ and $h$, and a positive correlation between $w$ and $A_s$, as predicted by our simple analysis~\footnote{To understand the sign of the correlation, notice that $w$ is negative, and so to make it larger in absolute value, it will move towards more negative values.}.

When adding the precise BAO data from the post-reconstructed power spectra, we are able to partially break the degeneracies.
A big improvement comes from adding small-$z$ BAO measurements,  as we have additional data points to break the $h-w$ degeneracy (and the $w-A_s$ one as a consequence), getting the final constraints in fig.~\ref{fig:wcdm_fs-allbao}. {The shift in the posteriors when adding small-$z$ BAO is consistent with what we observe in our tests on simulations, see App.~\ref{app:fsbao}. }

When we add SN data, we get a measurement of the luminosity distance $D_L = (1+z)^2 D_A$.
In the absence of calibration of absolute luminosities, we cannot determine $h$ and one gets the approximate degeneracy line $D_L(z=0.25) \sim \Omega_m^{-0.055} |w|^{0.1}$.
This is an anticorrelation between $\Omega_m$ and $w$, apparent from fig.~\ref{fig:wcdm_w-omega_m}.
As we discussed, FS+BAO data give an anticorrelation between $h$ and $w$ since $\omega_m$ is determined.
Therefore, $w$ and $\Omega_m \sim \omega_m / h^2$ are positively correlated.
This explains the good constraints we get when we add the SN dataset to FS+BAO.

\begin{table}
\centering
\scriptsize

\begin{tabular}{|p{3cm}|p{2cm}|p{2cm}|} 
 \hline 
$w$CDM & best-fit & mean$\pm\sigma$  \\ \hline 
$b_{1,{\rm CMASS \, NGC} }$ &$2.133$ & $2.434_{-0.34}^{+0.27}$  \\ 
$c_{2,{\rm CMASS \, NGC} }$ &$0.2843$ & $0.7325_{-1.1}^{+0.33}$ \\ 
$b_{1,{\rm CMASS \, SGC} }$ &$2.22$ & $2.492_{-0.32}^{+0.25}$  \\ 
$c_{2,{\rm CMASS \, SGC} }$ &$0.9084$ & $1.294_{-0.67}^{+0.57}$  \\ 
$b_{1,{\rm LOWZ \, NGC} }$ &$2.003$ & $2.242_{-0.26}^{+0.21}$  \\ 
$c_{2,{\rm LOWZ \, NGC} }$ &$0.7832$ & $1.283_{-0.77}^{+0.61}$ \\ 
\hline 
 \end{tabular}
 \begin{tabular}{|p{3cm}|p{2cm}|p{2cm}|} 
 \hline 
 $\nu \Lambda$CDM & best-fit & mean$\pm\sigma$  \\ \hline 
$b_{1,{\rm CMASS \, NGC} }$ &$2.049$ & $2.134_{-0.18}^{+0.16}$  \\ 
$c_{2,{\rm CMASS \, NGC} }$ &$0.4218$ & $0.8085_{-0.83}^{+0.28}$ \\ 
$b_{1,{\rm CMASS \, SGC} }$ &$2.143$ & $2.186_{-0.16}^{+0.15}$ \\ 
$c_{2,{\rm CMASS \, SGC} }$ &$1.172$ & $1.258_{-0.51}^{+0.47}$ \\ 
$b_{1,{\rm LOWZ \, NGC} }$ &$1.966$ & $2.019_{-0.15}^{+0.14}$  \\ 
$c_{2,{\rm LOWZ \, NGC} }$ &$0.9203$ & $1.273_{-0.58}^{+0.51}$ \\ 
\hline 
 \end{tabular}
\caption{\label{tab:eft_param} EFT parameters fitting BOSS FS + BOSS BAO on $w$CDM and $\nu \Lambda$CDM}
\end{table}

We finish this section with a comment on the EFT parameters measured in our analysis. 
Although our main focus in this paper is on constraining cosmological parameters in $w$CDM, the EFT parameters are also of interest as they are related to galaxy formation physics.
We show the 68\%-confidence intervals in Table~\ref{tab:eft_param} obtained fitting BOSS FS + BOSS BAO on $w$CDM, and on $\nu \Lambda$CDM for comparison. 
Details on the fit on $\nu \Lambda$CDM can be found in App.~\ref{app:fsbao}.
Only the EFT parameters that are actually varied in the fit are shown, while the others, as they appear only linearly in the galaxy power spectrum, are marginalized at the level of the likelihood, as performed in~\cite{DAmico:2019fhj,Colas:2019ret}. 
More details on the marginalization can be found in App.~\ref{app:marginalized}.
In Table~\ref{tab:eft_param}, we can see that the constraints on the EFT parameters are weaker in the fit on $w$CDM with respect to the one on $\nu \Lambda$CDM. 
In particular, the error bar on $b_1$ roughly doubles. 
This is expected from the discussion above on the broadband signal: 
varying $w$ opens larger degeneracies in the amplitude parameters, including the linear galaxy bias~$b_1$. 
Furthermore, the shift in $b_1$ going from $w$CDM to $\nu \Lambda$CDM is also consistent in direction and size with the correlation with $w$, that is found to be less than (but consistent with)~$-1$.
As the EFT parameters are clearly determined, this measurement could in principle allow us to learn on galaxy formation mechanisms. It would be interesting if a mapping between galaxy formation models and EFT parameters were to be performed.
Besides, given that the error bars on the EFT parameters are quite inflated in $w$CDM, doing this would also provide a good opportunity to use priors on the EFT parameters, motivated from simulations or more phenomenological models, in order to obtain more stringent constraints on the cosmological parameters. 
We leave these avenues for future work.

\section{PyBird: Python code for Biased tracers in redshift space}
\label{sec:pybird}

PyBird is a code written in Python 3, designed for evaluating the multipoles of the power spectrum of biased tracers in redshift space.
In general, PyBird can evaluate the power spectrum of matter or biased tracers in real or redshift space.
The equations on which PyBird is based can be found in~\cite{Perko:2016puo, DAmico:2019fhj}. 
The main technology used by the code is the FFTLog~\cite{Hamilton:1999uv}~\footnote{\href{https://jila.colorado.edu/~ajsh/FFTLog/index.html}{https://jila.colorado.edu/~ajsh/FFTLog/index.html}}, used to evaluate the one-loop power spectrum and the IR resummation, see sec.~\ref{sec:fastresum} for details. 

PyBird is designed for a fast evaluation of the power spectra, and can be easily inserted in a data analysis pipeline.
In fact, it is a standalone tool whose input is the linear matter power spectrum which can be obtained from any Boltzmann code, such as \code{CAMB}~\cite{Howlett:2012mh} or \code{CLASS}~\cite{Blas:2011rf}.
The PyBird output can be used in a likelihood code which can be part of the routine of a standard MCMC sampler.
The design is modular and concise, such that parts of the code can be easily adapted to other case uses (e.g., power spectrum at two loops or bispectrum).

The code is public and available at: \href{https://github.com/pierrexyz/pybird}{https://github.com/pierrexyz/pybird}, and it depends on the numerical libraries NumPy~\cite{Numpy} and SciPy~\cite{2020SciPy-NMeth}.
We also provide an explicit integration in the MCMC sampler MontePython 3~\cite{Brinckmann:2018cvx}, as well as a Jupyter notebook containing examples to start with.

PyBird can be used in different ways.
The code can evaluate the power spectrum either given one set of EFT parameters, or independently of the EFT parameters.
If the former option is faster, the latter is useful for subsampling or partial marginalization over the EFT parameters, or to Taylor expand around a fiducial cosmology for efficient parameter exploration~\cite{Colas:2019ret}. {PyBird runs in less than a second on a laptop.}

PyBird consists of the following classes:
\begin{itemize}
\item \code{Bird}: Main class which contains the power spectrum and correlation function, given a cosmology and a set of EFT parameters.
\item \code{Nonlinear}: given a \code{Bird()} object, computes the one-loop power spectrum and one-loop correlation function.
\item \code{Resum}: given a \code{Bird()} object, performs the IR-resummation of the power spectrum.
\item \code{Projection}: given a \code{Bird()} object, applies geometrical effects on the power spectrum: Alcock-Paczynski effect, window functions, fiber collisions, binning.
\item \code{Common}: containing shared objects among the other classes, such as $k$-array, multipole decomposition, etc.
\end{itemize}

We performed extensive tests on PyBird numerics, with particular attention on the numerical stability of the FFTLog.
Especially, zero padding and window are implemented following~\cite{McEwen:2016fjn}. 
All numerical parameters (e.g. number of points, boundaries or bias of the FFTLog's) are chosen such as, for a given value, we get the same power spectrum for that value multiplied or divided by 2, within $0.02\%$ for $l=0$, and $0.2\%$ for $l=2$, up to $k \sim 0.3$.
In the same spirit, the Taylor expansions in the IR-resummation are under control.

Notice finally that the particular implementation of the IR-resummation that we describe next allows for a simple modification of the code so that the dependence on $A_s$ can be factorized. Therefore, the sampling over~$A_s$ could be done extremely fast.

\subsection{Fast IR-resummation scheme}
\label{sec:fastresum}

The resummed power spectrum can be written as a sum of the nonresummed power spectrum plus IR-corrections. These IR-corrections can be efficiently evaluated using the FFTLog, which allows for a quick evaluation.
In this appendix, we derive the mathematical details of such implementation.

\begin{figure}[h]
\centering
\includegraphics[width=0.55\textwidth]{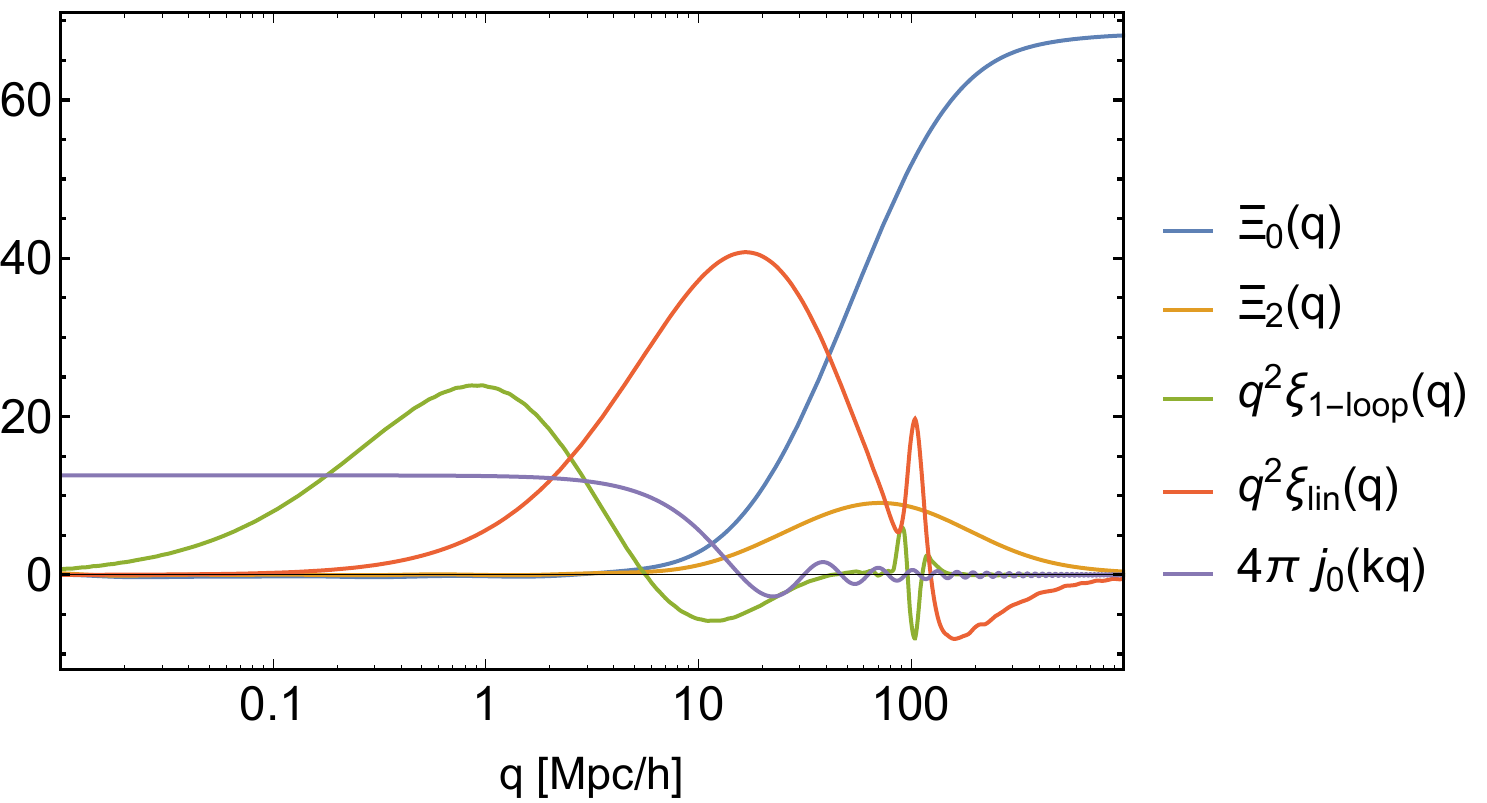}
\includegraphics[width=0.44\textwidth]{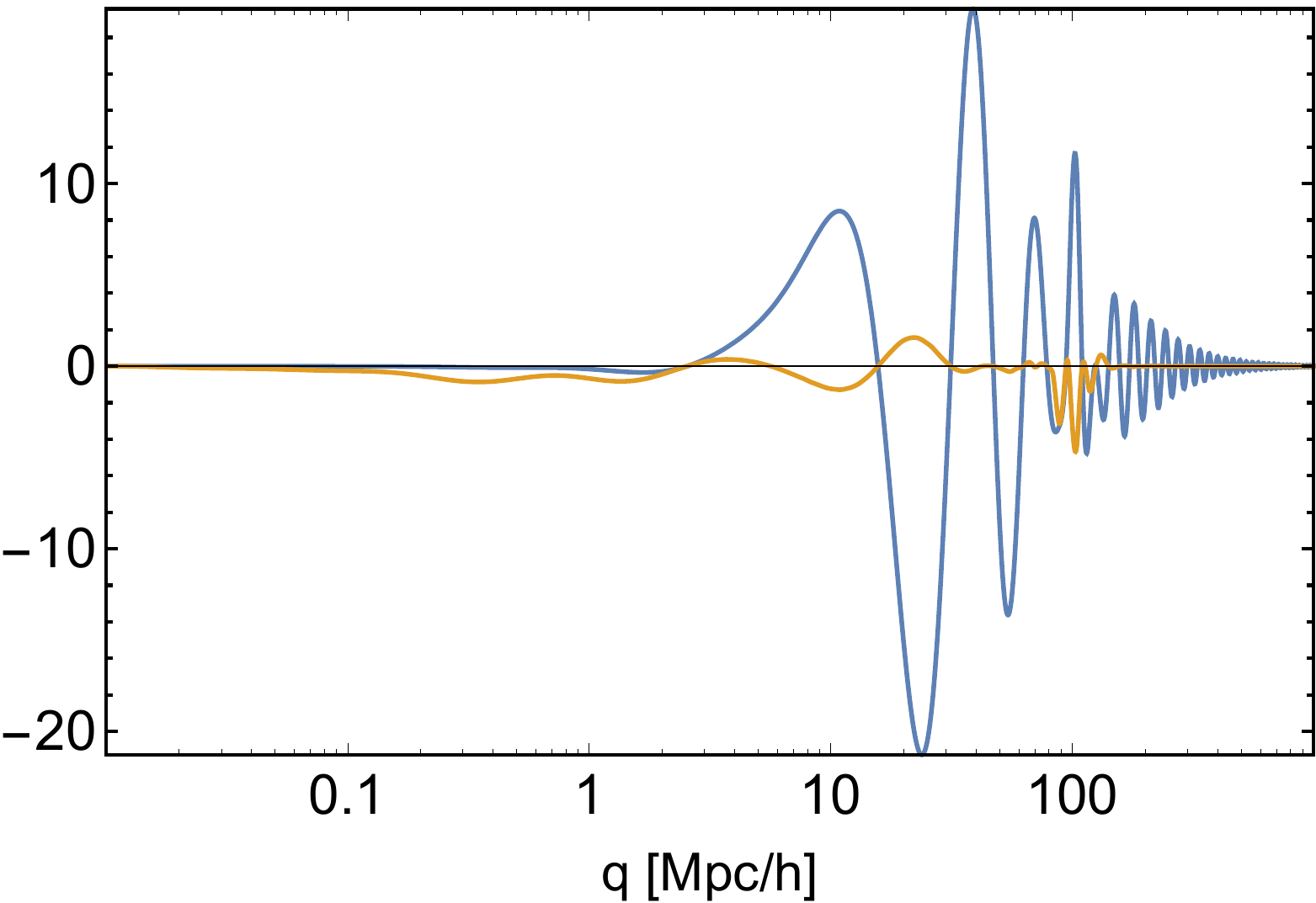}
\caption{ \label{fig:IRcorrections} {\it Left}: Various contributions appearing in the leading IR-corrections, Eq.~\eqref{eq:IRcorr} with $n=1$, for the matter real-space power spectrum, with $k = 0.2$. $\Xi_0(q)$ or $\Xi_2(q)$ are essentially acting as lowcut and a bandpass filters around the BAO peak, respectively, making the BAO the only relevant information for the IR-resummation, while the broadband is mostly filtered out. This illustrates the physical fact that the IR-resummation only acts on the wiggly part of the power spectrum.
{\it Right}: Typical IR-correction integrand for the linear part (blue) and the one-loop (orange), at $k= 0.2$, of the form: $4\pi q^2 \xi_j(q) \, [ e^{-\frac{k^2}{3} \Xi_0(q)} - 1 ] \, j_0(kq)$.  While the broadband information is irrelevant, most of the signal in the integral will come from the BAO. }
\end{figure}

After a straightforward manipulations, the IR-resummation in redshift space for biased tracers up to the $N$-loop order reads~\cite{Senatore:2014vja, Lewandowski:2015ziq}:
\begin{align}
P^\ell(k)|_N & = \sum_{j=0}^N \sum_{\ell'}  4\pi (-i)^{\ell'} \int dq \, q^ 2 \, Q_{||N-j}^{\ell \ell'}(k,q) \, \xi^{\ell'}_j (q), \label{eq:resumConvol}\\
\xi^{\ell'}_j (q) & = i^{\ell'}  \int  \frac{dp\, p^2}{2\pi^2} P^{\ell'}_j(p)  \, j_{\ell'}(pq).
\end{align}
where $|_N$ denotes the resummed power spectrum up to order $N$ and $P^{\ell}_j(k)$ and $\xi^{\ell}_j (k)$ are the $j$-loop order piece of the Eulerian power spectrum and correlation function, respectively. $Q_{||N-j}^{\ell \ell'}(k,q)$ encodes the effects from the bulk displacements and is given by:
\begin{align}
Q_{||N-j}^{\ell \ell'}(k,q) & = \frac{2\ell+1}{2} \int_{-1}^{1}d\mu_k \,\frac{i^{\ell'}}{4 \pi} \int d^2 \hat{q} \, e^{-i\q \cdot \k} \, F||_{N-j}(\k,\q) \mathcal{P}_\ell(\mu_k) \mathcal{P}_{\ell'}(\mu_q), \label{eq:resumQ}\\
F||_{N-j}(\k,\q) & = T_{0,r}(\k,\q)\times T_{0,r}^{-1}||_{N-j}(\k,\q), \nonumber\\
T_{0,r}(\k,\q) & = \exp \left\lbrace -\frac{k^2}{2} \left[ \Xi_0(q) (1+2 f \mu_k^2 + f^2 \mu_k^2) + \Xi_2(q) \left( (\hat k \cdot \hat q)^2 + 2 f \mu_k \mu_q (\hat k \cdot \hat q) + f^2 \mu_k^2 \mu_q^2 \right) \right] \right\rbrace , \nonumber
\end{align}
where $\Xi_0(q)$ and $\Xi_2(q)$ are given by:
\begin{align}
\Xi_0(q) & = \frac{2}{3} \int \frac{dp}{2\pi^2} \, \exp \left(-\frac{p^2}{\Lambda_{\rm IR}^2} \right)  P_{11}(p) \, \left[1 - j_0(pq) - j_2(pq) \right],\\
\Xi_2(q) & = 2 \int \frac{dp}{2\pi^2}\, \exp \left( -\frac{p^2}{\Lambda_{\rm IR}^2} \right)  P_{11}(p) \, j_2(pq) .
\end{align}
By expanding the exponential {in $F||_{N-j}(\k,\q)$ in powers of $k^2$} and performing the angular integrals in Eq.~\eqref{eq:resumQ} \footnote{To perform the angular integrals, we follow the steps of \cite{Lewandowski:2015ziq}, but here expanding the whole argument of the exponential.}, the terms in Eq.~\eqref{eq:resumConvol} can be put in the form of:
\begin{equation}\label{eq:IRcorr}
4\pi (-i)^{\ell'} k^{2n} \, \mathcal{Q}_{||N-j}^{\ell \ell'}(n,\alpha) \, \int dq \, q^ 2 \,  \left[ \Xi_i(q) \right]^n   \xi_j^{\ell'}(q) \, j_{\alpha}(kq),
\end{equation}
where $n$ is the integer power controlling the expansion of the exponential, $j_{\alpha}$ is the $\alpha$-th order spherical Bessel function, $\left[ \Xi_i(q) \right]^n$ denotes a product of the form $\Xi_0(q) \times ...  \times \Xi_0(q) \times \Xi_2(q) \times ... \times \Xi_2(q)$ such that the total number of terms in the product is $n$, and $\mathcal{Q}_{||N-j}^{\ell \ell'}(n, \alpha)$ is a number that depends on $N-j$, $\ell$, $\ell'$, $n$, $\alpha$ (and $f$).
In particular, at lowest order $n=0$, one gets nothing but the nonresummed power spectrum.
The contributions to the integrand of Eq.~\eqref{eq:IRcorr} are shown in Fig.~\ref{fig:IRcorrections}. 

Thus, the resummed power spectrum is equal to the nonresummed power spectrum plus IR-corrections written as an expansion in powers of $k^2$ (starting at $k^2$):
\begin{equation}\label{eq:masterIRformula}
P^\ell(k)|_N = P^\ell(k) + \sum_{j=0}^N \sum_{\ell'} \sum_{n=1} \sum_{\alpha} 4\pi (-i)^{\ell'} k^{2n} \,\mathcal{Q}_{||N-j}^{\ell \ell'}(n, \alpha) \, \int dq \, q^ 2 \,  \left[ \Xi_i(q) \right]^n   \xi_j^{\ell'}(q) \, j_{\alpha}(kq) \, ,
\end{equation}
{where we remind that $n$ is the integer controlling the expansion in powers of $k^2$ of the exponential of the bulk displacements, running up to a sufficient order to achieve convergence of the expansion up to the highest $k$ mode of interest. The explicit expressions for $\mathcal{Q}_{||N-j}^{\ell \ell'}(n, \alpha)$ can be found in PyBird repo either in the main code or in the Mathematica notebook. }
Therefore, evaluating the resummed power spectrum boils down to evaluating one-dimensional integrals that can be performed using the FFTLog.
The loops of the (nonresummed) power spectrum and correlation function can be evaluated using the FFTLog, see e.g.~\cite{Simonovic:2017mhp, Lewandowski:2018ywf}.
Then, the IR-resummation consists simply in correcting the power spectrum with a set of spherical Bessel transforms that, once again, can be performed using the FFTLog.
In practice, we find that expanding up to $n=8$ is sufficient to achieve convergence up to $k\sim 0.3 \hinvMpc$ for both $\ell = 0, 2$, and keeping terms contributing only significantly\footnote{Precisely, we keep only the terms with one power in $k^2 \Xi_2(q)$ (see \cite{Lewandowski:2015ziq} for the validity of such approximation), and drop a few more terms that are found to be irrelevant.}, the IR-resummation can be achieved with about $30$ FFTLog's. 
Notice that the spherical Bessel transforms in Eq.~\ref{eq:masterIRformula} all present well-localized compact integrands around the BAO scales, allowing to perform the FFTLog with a few points ($\sim \mathcal{O}(200)$) to achieve good accuracy. 

Finally, as discussed in ~\cite{Senatore:2014vja, Lewandowski:2015ziq}, the dependence on the displacements is analytic, and therefore the IR-resummation will agree to perturbation theory once this one is performed to extremely high order. In practice, indeed, in our procedure we are Taylor expanding the IR-resummation, which is equivalent to including the part of the perturbative loop that would encode the effect of the long displacements. The fact that we go to order $n=8$ means that effectively we are doing an eight-loop calculation, but effectively keeping track only of the part that is relevant for the IR-resummation.

\section*{Acknowledgements}
We thank Florian Beutler and H\'ector Gil-Mar\'in for support with the BOSS data, and Emanuele Castorina for useful discussions.
LS and GDA are partially supported by Simons Foundation Origins of the Universe program (Modern Inflationary Cosmology collaboration) and LS by NSF award 1720397. 
GDA and PZ thank the Theoretical Physics Group at CERN for hospitality during part of this work.
PZ thanks Fran\c{c}oise Papa for hospitality during completion of this work.
The linear power spectra were computed with the CLASS Boltzmann code~\cite{Blas_2011}~\footnote{ \href{http://class-code.net}{http://class-code.net}}. 
The posteriors were sampled using MontePython cosmological parameter inference code~\cite{Brinckmann:2018cvx, Audren:2012wb}~\footnote{ \href{https://github.com/brinckmann/montepython\_public}{https://github.com/brinckmann/montepython\_public}}.
 Part of the analysis was performed on the Sherlock cluster at the Stanford University, for which we thank the support team, and part on the HPC (High Performance Computing) facility of the University of Parma, whose support team we thank.

\appendix

\section{FS + BAO joint analysis}\label{app:fsbao}

The post-reconstructed power spectrum contains additional information in the BAO with respect to the pre-reconstructed one, by adding displacements from higher (pre-reconstructed) $n$-point functions. Here we describe how we analyse the BOSS DR12 pre- and post-reconstructed power spectrum.

\subsection{FS analysis}
The full shape (FS) of the pre-reconstructed power spectrum is fit with a cosmology- and survey-dependent theoretical model.
For all analyses presented in this work, we fit the FS using the EFTofLSS galaxy power spectrum with IR-resummation, Alcock-Paczynski effect, window function~\footnote{The window function has been remeasured using the technique of~\cite{Beutler:2018vpe}. We thank Florian Beutler for providing them to us in~\cite{DAmico:2019fhj,Colas:2019ret}.}, correction for fiber collisions, and priors as described in~\cite{DAmico:2019fhj}. 
Note that when analyzing extension to the concordance model such as $w$CDM, the growth rate $f$ and the relations implied in the Alcock-Paczynski effect --- through $H(z)$ and $D_A(z)$ --- are modified accordingly. Note that this analysis is very similar operationally to the analyses that are normally carried out for the CMB, {\it i.e.} all the data points are fit with a theoretical model that is updated at every cosmology ({\it CMB style}).  

Let us summarize the EFT model for the FS. 
The galaxy power spectrum at one loop in redshift space reads~\cite{Perko:2016puo}:
\begin{align}\label{eq:powerspectrum}
& P(k, \mu) = Z_1(\mu)^2 P_{11}(k) \nonumber  \\
& + 2 \int \frac{d^3q}{(2\pi)^3}\; Z_2(\q,\k-\q,\mu)^2 P_{11}(|\k-\q|)P_{11}(q) + 6 Z_1(\mu) P_{11}(k) \int\, \frac{d^3 q}{(2\pi)^3}\; Z_3(\q,-\q,\k,\mu) P_{11}(q)\nonumber \\
& + 2 Z_1(\mu) P_{11}(k)\left( c_\text{ct}\frac{k^2}{{ k^2_\textsc{m}}} + c_{r,1}\mu^2 \frac{k^2}{k^2_\textsc{m}} + c_{r,2}\mu^4 \frac{k^2}{k^2_\textsc{m}} \right) + \frac{1}{\bar{n}_g}\left( c_{\epsilon,0}+c_{\epsilon,1}\frac{k^2}{k_\textsc{m}^2} + c_{\epsilon,2} f\mu^2 \frac{k^2}{k_\textsc{m}^2} \right) \, ,
\end{align}
where $P_{11}$ denotes the linear matter power spectrum, $\mu$ is the component in the direction of the line-of-sight of the wavenumber $\k$ , $k^{-1}_\textsc{m} (\simeq k^{-1}_\textsc{nl})$ is the scale controlling the bias (dark matter) derivative expansion, and $\bar{n}_g$ is the mean galaxy number density.
Here $Z_1,Z_2$ and $Z_3$ are the redshift-space galaxy density kernels, given by:
\begin{align}\label{eq:redshift_kernels}\nonumber
    Z_1(\q_1) & = K_1(\q_1) +f\mu_1^2 G_1(\q_1) = b_1 + f\mu_1^2,\\ \nonumber
    Z_2(\q_1,\q_2,\mu) & = K_2(\q_1,\q_2) +f\mu_{12}^2 G_2(\q_1,\q_2)+ \, \frac{1}{2}f \mu q \left( \frac{\mu_2}{q_2}G_1(\q_2) Z_1(\q_1) + \text{perm.} \right),\\ \nonumber
    Z_3(\q_1,\q_2,\q_3,\mu) & = K_3(\q_1,\q_2,\q_3) + f\mu_{123}^2 G_3(\q_1,\q_2,\q_3) \nonumber \\ 
    &+ \frac{1}{3}f\mu q \left(\frac{\mu_3}{q_3} G_1(\q_3) Z_2(\q_1,\q_2,\mu_{123}) +\frac{\mu_{23}}{q_{23}}G_2(\q_2,\q_3)Z_1(\q_1)+ \text{cyc.}\right),
\end{align}
where $\mu= \q \cdot \hat{\z}/q$, $\q = \q_1 + \dots +\q_n$, and $\mu_{i_1\ldots  i_n} = \q_{i_1\ldots  i_n} \cdot \hat{\z}/q_{i_1\ldots  i_n}$, $\q_{i_1 \dots i_m}=\q_{i_1} + \dots +\q_{i_m}$, with $\hat{\z}$ being the line of sight unit vector and $n$ is the order of the kernel $Z_n$. 
The galaxy density kernels $K_n$ are given in the basis of descendants by~\cite{Angulo:2015eqa,Fujita:2016dne}:
 \begin{align}
     K_1 & = b_1, \\
     K_2(\q_1,\q_2) & = b_1\frac{\q_1\cdot \q_2}{q_1^2}+ b_2\left( F_2(\q_1,\q_2)- \frac{\q_1\cdot \q_2}{q_1^2} \right) + b_4 + \text{perm.} \, , \\
     K_3(k, q) & = \frac{b_1}{504 k^3 q^3}\left( -38 k^5q + 48 k^3 q^3 - 18 kq^5 + 9 (k^2-q^2)^3\log \left[\frac{k-q}{k+q}\right] \right) \\
    &+ \frac{b_3}{756 k^3 q^5} \left( 2kq(k^2+q^2)(3k^4-14k^2q^2+3q^4)+3(k^2-q^2)^4 \log \left[\frac{k-q}{k+q}\right] \nonumber \right) \, ,
 \end{align}
where $F_2$ is the second order density kernel in standard perturbation theory, and the velocity kernels $G_n$ are simply the standard perturbation theory ones (see~e.g.~\cite{Bernardeau:2001qr} for explicit expressions). 
Here the third-order kernel is given after performing the angular integration over $x = \hat k \cdot \hat q$ and UV-subtraction. 
Following~\cite{DAmico:2019fhj}, we restrict the EFT parameters to vary only within their physical range, imposing a Gaussian prior centered on $0$ of size $2$ on $b_3$, $c_{\rm ct}$, $c_{\epsilon,0}/\bar n_g$, and $c_{\epsilon, {\rm quad}}/\bar n_g \equiv \tfrac{2}{3} c_{\epsilon, 2} f /\bar n_g$, and of size 4 on $c_{r,1}$ and $c_{r,2}$; and a flat prior on $b_1$ and $c_2 \equiv (b_2+b_4)/\sqrt{2}$ of $[0, 4]$ and $[-4, 4]$, respectively. 
We have also set $b_2-b_4 = 0$ and $c_{\epsilon, {\rm mono}}/\bar n_g \equiv (c_{\epsilon,1} + \tfrac{1}{3}c_{\epsilon,2}f)/\bar n_g = 0$, since we find that the signal-to-noise ratio of the data is such that we cannot measure those combinations of parameters.
We use one set of EFT parameters per skycut.

We expand in multipoles the resulting power spectrum and we apply the IR-resummation as in~(\ref{eq:resumConvol}).
Then, we apply the the Alcock-Paczynski transformation, which comes from the fact that the estimation of galaxy spectra assumes a reference cosmology to transform redshift and celestial coordinates into Cartesian coordinates.
The wavenumbers parallel and perpendicular to the line of sight ($k_\parallel$, $k_\perp$) are related to the ones of the reference cosmology as ($k^{\rm ref}_\parallel$, $k^{\rm ref}_\perp$) as:
\begin{equation}
    k^{\rm ref}_\parallel = q_{\parallel} k_\parallel \, , \qquad
    k^{\rm ref}_\perp = q_{\perp} k_\perp \, ,
\end{equation}
where the distortion parameters are defined by
\begin{equation}
    q_{\parallel} = \frac{D_A(z) H(z=0)}{D_A^{\rm ref}(z) H^{\rm ref}(z=0)} \, , \qquad
    q_{\perp} = \frac{H^{\rm ref}(z) / H^{\rm ref}(z=0)}{H(z) / H(z=0)} \, ,
\end{equation}
where $D_A$, $H$ are the angular diameter distance and Hubble parameter, respectively, and `$\textrm{ref}$' denotes quantities calculated in the reference cosmology.
In terms of these parameters, the power spectrum multipoles in the reference cosmology is given by
\begin{equation}
    P_{\ell}(k) = \frac{2 \ell + 1}{2 q_\parallel q_\perp^2} \int_{-1}^{1} \mathrm{d} \mu^{\rm ref} \, P(k(k^{\rm ref}, \mu^{\rm ref}), \mu(\mu^{\rm ref})) \,
    \mathcal{L}_\ell(\mu^{\rm ref}) \, ,
\end{equation}
where we have
\begin{equation}
    k = \frac{k^{\rm ref}}{q_\perp} \left[ 1 + (\mu^{\rm ref})^2 \left( \frac{1}{F^2} -1 \right) \right]^{1/2} \, , \qquad \mu = \frac{\mu^{\rm ref}}{F} \left[ 1 + (\mu^{\rm ref})^2 \left( \frac{1}{F^2} -1 \right) \right]^{-1/2} \, ,
\end{equation}
with $F=q_\parallel / q_\perp$.
Finally, we convolve with the window functions and correct for fiber collisions as discussed in~\cite{DAmico:2019fhj}.

\subsection{Reconstructed BAO} For the post-reconstructed power spectra, we follow the standard treatment with fixed template but varying BAO parameters~\cite{Gil-Marin:2015nqa, Beutler:2016ixs}.

For each BOSS skycut, we fit the reconstructed power spectrum with the template described in~\cite{Beutler:2016ixs}.
In addition to nuisance parameters, the fit measures the following two geometrical distortion parameters:
\begin{equation}
\alpha_\parallel = \frac{H^{\rm ref}(z) r_s^{\rm ref}(z_d)}{H(z) r_s(z_d)} \, , \qquad
\alpha_\perp = \frac{D_A(z) r_s^{\rm ref}(z_d)}{D_A^{\rm ref}(z) r_s(z_d)} \, ,
\end{equation}
where $r_s(z_d)$ is the sound horizon at the drag epoch.
The $\alpha_\parallel$, $\alpha_\perp$ parameters contain the information from the reconstructed BAO.

\subsection{FS+BAO joint analysis} When combining FS and BAO, there will be a sizeable covariance between the BAO parameters and the pre-reconstructed power spectra, which we measure from 2048 patchy mocks as follows. 
For the full shape part, we estimate the covariance from the Patchy pre-reconstructed power spectrum measurements.
For the BAO parameters, we determine the 2048 bestfit points from the Patchy post-reconstructed power spectrum measurements.
The joint covariance is then estimated, together with the cross-correlation between the pre-reconstructed Patchy power spectra and the BAO parameters. 
Explicitly, the joint covariance is estimated as:
\begin{equation}
C^{ij} = \frac{1}{N_m-1} \sum_{n=1}^{N_m} \left( V^{i}_n - \bar{V}^{i} \right) \left( V^{j}_n - \bar{V}^{j} \right)  \, ,
\end{equation}
where $N_m$ is the number of mocks. 
Here $i$ or $j$ are indices running over the pre-reconstructed power spectrum multipoles and BAO parameters of each mock $n$ represented by the vector $\mathbf{V}_n \equiv \lbrace P_0(k_0), \dots P_0(k_N), P_2(k_0), \dots, P_2(k_N), \alpha_\parallel, \alpha_\perp \rbrace_n$, where $N$ is the number of $k$-bins, and:
\begin{equation}
\bar{V}^i = \frac{1}{N_m}  \sum_{n=1}^{N_m} V^{i}_n \, 
\end{equation}
is the mean over all the mocks.

Given the covariance, the data and the EFT model, we define a Gaussian likelihood $\mathcal{L}$ of the data given the cosmological and EFT parameters:
\begin{equation}
    \ln \mathcal{L} = - \frac{1}{2} \sum_{\al, \bt} (P^{\rm EFT}_\al - D_\al) C^{-1}_{\al \bt} (P^{\rm EFT}_\bt - D_\bt) \, ,
\end{equation}
where we use a concise notation in which $P$ is the vector of the EFT power spectra and the $(\al_{\parallel}$, $\al_\perp$) parameters for each $k$-bin, multipole moment and skycut, $D$ the corresponding vector of the data, and $C$ the covariance matrix constructed as a block-diagonal matrix for each skycut, with each block calculated as discussed before.

\begin{figure}[h]
\centering
\includegraphics[width=0.76\textwidth]{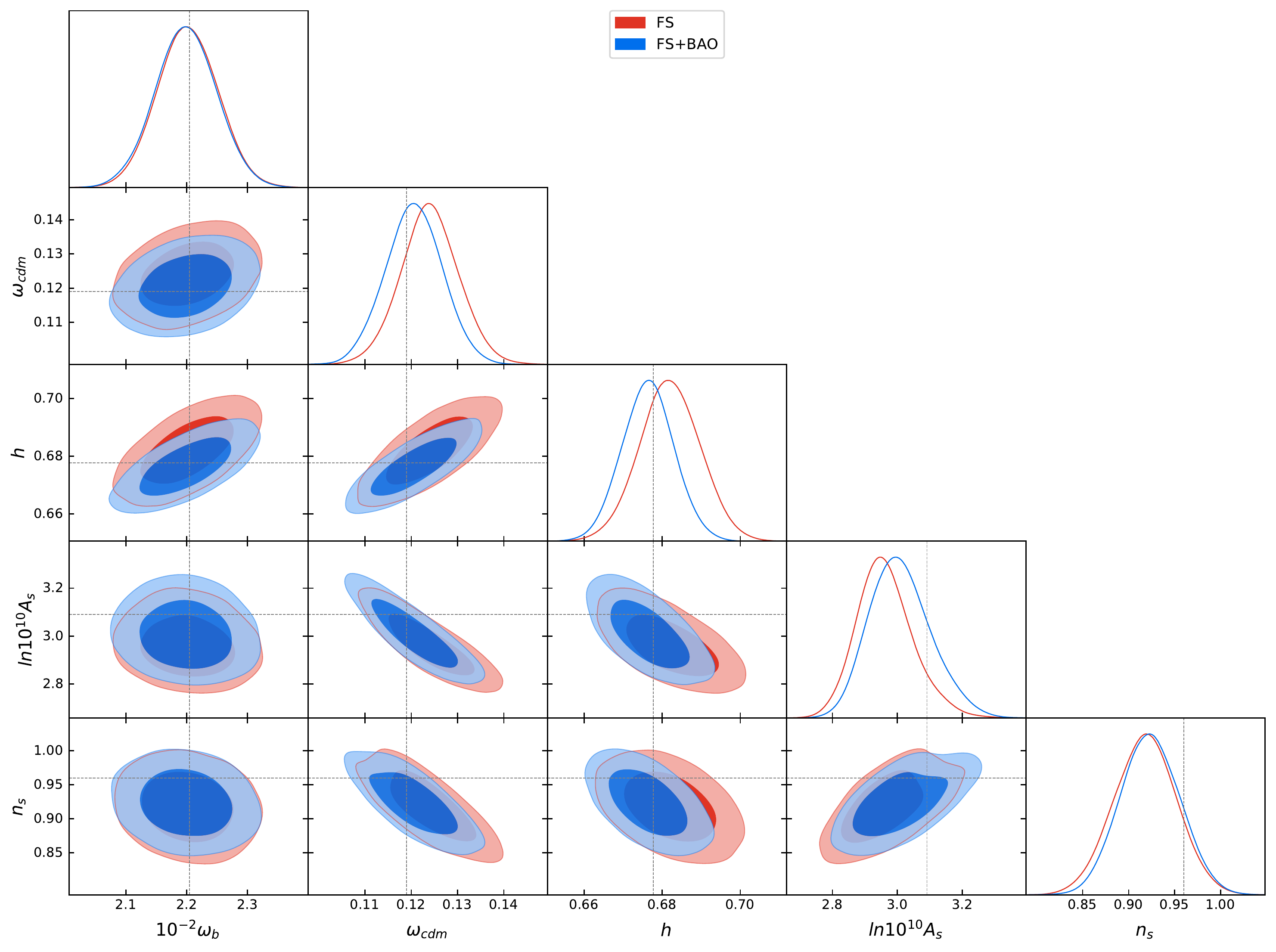}
\caption{ \label{fig:patchy} Triangle plot of $\Lambda$CDM fitting the mean of 2048 patchy CMASS NGC mocks FS or FS + BAO, using a covariance rescaled by 16 and a BBN prior on $\omega_b$.
The dashed lines represent the truth of the simulations.}
\end{figure}

The FS analysis has been already extensively checked on $\Lambda$CDM against simulations in~\cite{DAmico:2019fhj, Colas:2019ret}: the theory-systematic error is under control at less than $\sigma_{\rm stat}/4$, where $\sigma_{\rm stat}$ is the statistical error obtained by fitting the BOSS DR12 pre-reconstructed power spectra up to $k_{\rm max} \sim 0.2 \hinvMpc$. 
For the joint FS+BAO analysis, we perform the following tests:

\paragraph{Test against simulations}
In Fig.~\ref{fig:patchy} and Table~\ref{tab:patchy}, we show the marginalized posterior distribution of the $\Lambda$CDM parameters with a BBN prior on $\omega_b$, obtained by fitting the mean over 2048 CMASS NGC Patchy mocks, analyzed with the covariance divided by 16, as done in the tests performed in~\cite{DAmico:2019fhj, Colas:2019ret}.
This rescaling allows us to measure the theory-systematic error within about $1/3 \, \sigma_{\rm stat}$ of the individual 1D posteriors, where $\sigma_{\rm stat}$ is the $68\%$ confidence interval obtained by fitting BOSS data.
By measuring the theory-systematic error on simulations as the distance of the $68\%$ confidence interval of the 1D posterior to the Patchy true value, we find no theory-systematic error except a tiny one on $n_s$ of less than $0.01$ for both FS and FS+BAO, and on $\ln (10^{10}A_s)$ of at most $1/4 \,\sigma_{\rm stat}$ for FS and less for FS+BAO, thus safely negligible.
Importantly, adding BAO moves very little the mean of the posteriors: the shifts are at most $1/4 \,\sigma_{\rm stat}$.
This is already safe when analyzing BOSS data alone, and the systematic error becomes completely negligible when combined with the Planck2018 likelihood.
Notice also that adding BAO moves the posteriors towards the true cosmology of the simulation.

\begin{table}
 \centering
\scriptsize
 \begin{tabular}{|l|c|c|c|} 
 \hline 
FS & best-fit & mean$\pm\sigma$ & truth  \\ \hline 
$\omega_{cdm }$ &$0.1231$ & $0.124_{-0.0064}^{+0.0064}$ & 0.119\\ 
$h$ &$0.6816$ & $0.682_{-0.0079}^{+0.008}$ & 0.6777\\ 
$\ln\left(10^{10}A_{s }\right)$ &$2.95$ & $2.963_{-0.096}^{+0.076}$ & 3.091\\ 
$n_{s }$ &$0.9377$ & $0.9181_{-0.035}^{+0.035}$ & 0.96 \\ 
\hline 
 \end{tabular} 
 \begin{tabular}{|l|c|c|c|} 
 \hline 
FS+BAO & best-fit & mean$\pm\sigma$ & truth\\ \hline 
$\omega_{cdm }$ &$0.1179$ & $0.1205_{-0.0062}^{+0.0063}$ & 0.119\\ 
$h$ &$0.6744$ & $0.6763_{-0.0068}^{+0.0067}$ & 0.6777 \\ 
$\ln\left(10^{10}A_{s }\right)$ &$3.057$ & $3.01_{-0.11}^{+0.084}$ & 3.091\\ 
$n_{s }$ &$0.9205$ & $0.9236_{-0.034}^{+0.033}$  & 0.96 \\ 
\hline 
 \end{tabular} \\ 
 \caption{\label{tab:patchy} Results on $\Lambda$CDM fitting the mean of 2048 patchy CMASS NGC mocks FS or FS + BAO, using a covariance rescaled by 16 and a BBN prior on $\omega_b$.    }
\end{table}

\begin{figure}[h]
\centering
\includegraphics[width=0.76\textwidth]{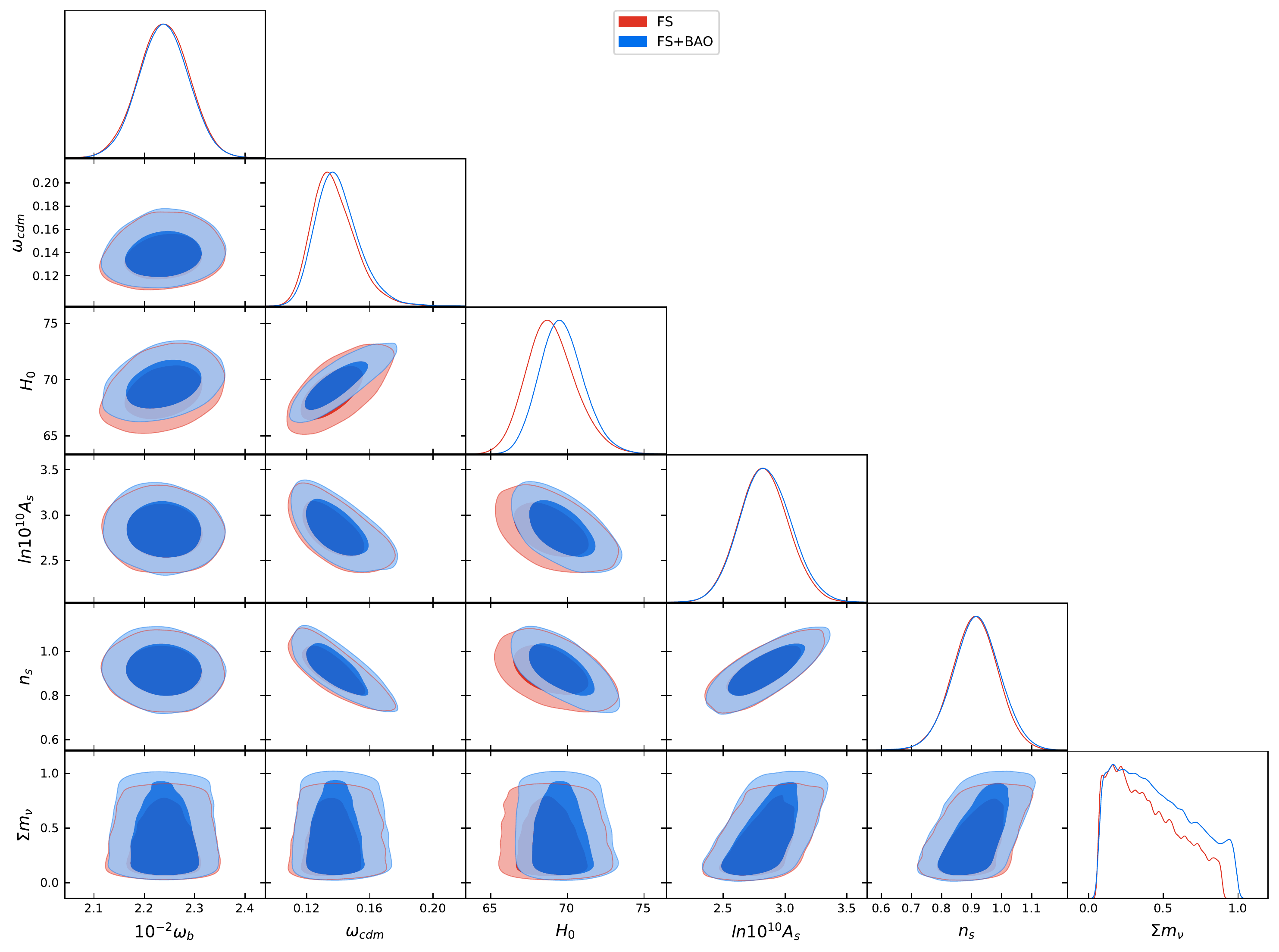}
 \caption{\label{fig:ncdm_eft_bao} Triangle plot of $\Lambda$CDM + massive neutrinos fitting BOSS FS and BOSS FS + BAO with a BBN prior on $\omega_b$. }
\end{figure}

\paragraph{Fit to BOSS data}
We fit the BOSS data with a $\nu \Lambda$CDM model with 3 massive neutrinos obeying a normal hierarchy with a flat prior on the total mass: $0.06 \leq \sum_i m_{\nu, i} /\mathrm{eV} \leq 1.0$ and a BBN prior on $\omega_b$. The results are shown in Fig.~\ref{fig:ncdm_eft_bao} and in Table~\ref{tab:ncdm_eft_bao}. We find no shift in the cosmological parameters when adding BAO except on $H_0$, for which the shift is less that $1/4 \,\sigma_{\rm stat}$.

\begin{table}
\centering
\scriptsize

\begin{tabular}{|l|c|c|} 
 \hline 
FS & best-fit & mean$\pm\sigma$ \\ \hline 
$\omega_{cdm }$ &$0.1294$ & $0.1369_{-0.016}^{+0.011}$ \\ 
$H_0$ &$68.96$ & $68.88_{-1.8}^{+1.4}$ \\ 
$\ln\left(10^{10}A_{s }\right)$ &$2.871$ & $2.834_{-0.21}^{+0.2}$ \\ 
$n_{s }$ &$0.908$ & $0.9134_{-0.074}^{+0.077}$ \\ 
$\Sigma m_\nu $ &$0.1702$ & $0.3901_{-0.33}^{+0.11}$  \\  \hline 
$\sigma_8$ &$0.7484$ & $0.7217_{-0.05}^{+0.045}$ \\ 
\hline 
 \end{tabular} 
\begin{tabular}{|l|c|c|} 
 \hline 
FS+BAO & best-fit & mean$\pm\sigma$  \\ \hline 
$\omega_{cdm }$ &$0.1315$ & $0.1398_{-0.016}^{+0.011}$  \\ 
$H_0$ &$68.94$ & $69.66_{-1.6}^{+1.3}$  \\ 
$\ln\left(10^{10}A_{s }\right)$ &$2.852$ & $2.84_{-0.21}^{+0.2}$  \\ 
$n_{s }$ &$0.901$ & $0.9153_{-0.08}^{+0.08}$  \\ 
$\Sigma m_\nu$ &$0.2483$ & $0.4526_{-0.39}^{+0.12}$  \\ \hline 
$\sigma_8$ &$0.733$ & $0.7258_{-0.05}^{+0.045}$  \\ 
\hline 
 \end{tabular} \\ 
 \caption{\label{tab:ncdm_eft_bao} Results on $\Lambda$CDM + massive neutrinos fitting BOSS FS (+ BAO) with a BBN prior on $\omega_b$. }
 \end{table}

\paragraph{Covariance}  Given that the the pre-reconstruction and the post-reconstraction data set are quite correlated, it is important to accurately measure the cross correlation in order to ensure potential cancellations. 
We compare the covariance measured from 2048 patchy mocks against another covariance measured from 1024 Patchy mocks. We find the same amount of correlation between the BAO parameters, as well as between them and the individual $k$-bins of the FS. We analyze the Patchy mocks as in Fig.~\ref{fig:patchy} but with the covariance measured with half of the mocks and find perfect match between the two results. {This check ensures that our measurements of the covariance are accurate enough.}

\paragraph{Combination with Planck data}
Combining Planck2018 with the BAO parameters we measured leads to the same results as Planck2018 combined with the BOSS DR12 consensus measurements \cite{Aghanim:2018eyx}.

\section{Theory-systematic error} \label{app:theoryerror}

As mentioned in App.~\ref{app:fsbao},  the EFTofLSS has already being extensively calibrated against simulations (see e.g. \cite{DAmico:2019fhj,Colas:2019ret,Nishimichi:2020tvu}):
it was found that the BOSS FS can be analyzed up to $\kmax = 0.23 \hinvMpc$~\cite{Colas:2019ret}. 
In App.~\ref{app:fsbao}, we have also seen that this remains valid upon the addition of the BOSS BAO measured from the post-reconstructed power spectrum. 
These conclusions were reached for $\Lambda$CDM with a BBN prior. 
We now perform the same tests but for $w$CDM with a BBN prior.
We first focus on the patchy mocks as it allows us to test for the joint FS+BAO analysis, and then present the results using N-body simulations.

\begin{figure}[h]
\centering
\includegraphics[width=0.76\textwidth]{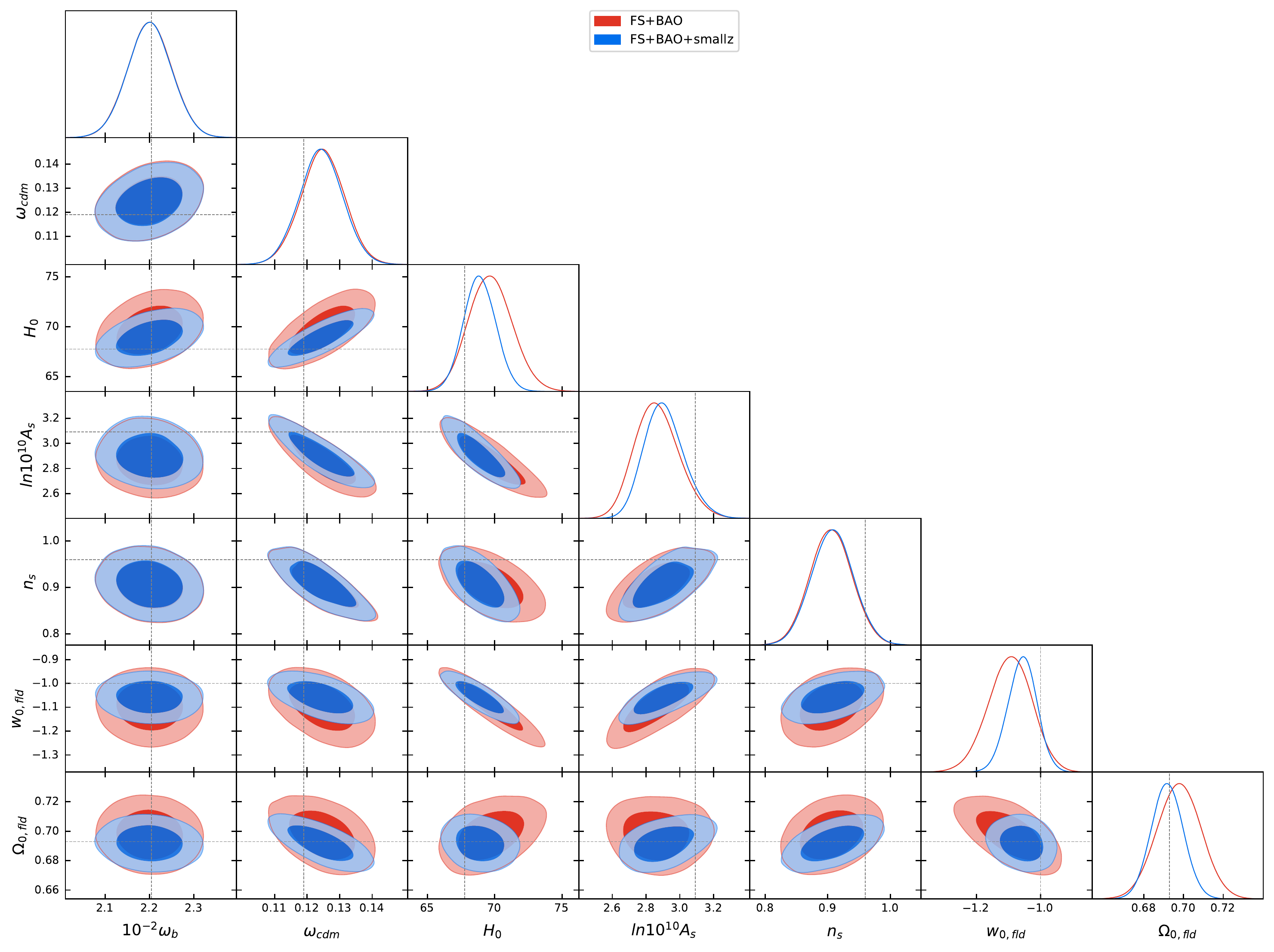}
\caption{ \label{fig:patchy_wcdm} Triangle plot of $w$CDM fitting the mean of 2048 patchy CMASS NGC mocks FS + BAO with and without `small-$z$' BAO prior., using a covariance rescaled by 16 and a BBN prior on $\omega_b$.
The dashed lines represent the truth of the simulations.}
\end{figure}

\begin{table}
\centering
\scriptsize
\begin{tabular}{|p{1.4cm}|c|c|c|p{1.4cm}|} 
 \hline 
{\tiny FS+BAO w/o small-$z$} & best-fit & mean$\pm\sigma$ & truth &theory sys. in $\sigma_{\rm data}$\\ \hline 
$\omega_{cdm }$ &$0.1258$ & $0.1248_{-0.0067}^{+0.007}$ & $0.119$ & $0\%$ \\ 
$H_0$ &$69.6$ & $69.69_{-1.7}^{+1.6}$  & $67.77$ & $5\%$ \\ 
$\ln\left(10^{10}A_{s }\right)$ &$2.841$ & $2.867_{-0.15}^{+0.12}$ & $3.091$ & $35\%$\\ 
$n_{s }$ &$0.9116$ & $0.9057_{-0.035}^{+0.034}$ & $0.96$ & $25\%$ \\ 
$w_{0,fld }$ &$-1.086$ & $-1.095_{-0.066}^{+0.072}$ &$-1$ & $10\%$ \\ 
$\Omega_{0,fld }$ &$0.695$ & $0.6977_{-0.011}^{+0.012}$ & $0.693$ & $0\%$ \\ 
\hline 
 \end{tabular}
 \begin{tabular}{|p{1.4cm}|c|c|c|p{1.4cm}|} 
 \hline 
{\tiny FS+BAO w/ small-$z$} & best-fit & mean$\pm\sigma$ & truth & theory sys. in $\sigma_{\rm data}$\\ \hline  
$\omega_{cdm }$ &$0.1248$ & $0.1243_{-0.0068}^{+0.0068}$ & $0.119$ & $0\%$ \\ 
$H_0$ &$68.77$ & $68.88_{-1.3}^{+1.2}$ & $67.77$ & $0\%$ \\ 
$\ln\left(10^{10}A_{s }\right)$ &$2.873$ & $2.908_{-0.13}^{+0.11}$ & $3.091$ & $8\%$ \\ 
$n_{s }$ &$0.9175$ & $0.9077_{-0.035}^{+0.034}$ & $0.96$ & $2\%$ \\ 
$w_{0,fld }$ &$-1.053$ & $-1.057_{-0.046}^{+0.048}$ & $-1$ & $1\%$ \\ 
$\Omega_{0,fld }$ &$0.6897$ & $0.6918_{-0.0084}^{+0.0083}$ & $0.693$ & $0\%$ \\ 
\hline 
 \end{tabular} \\
 \caption{\label{tab:patchy_wcdm} Results on $w$CDM fitting the mean of 2048 patchy CMASS NGC mocks FS + BAO with and without `small-$z$' BAO prior, using a covariance rescaled by 16 and a BBN prior on $\omega_b$. The theory-systematic errors are quoted in percentage of the corresponding $68\%$ confidence intervals, $\sigma_{\rm data}$, obtained analyzing FS+BAO (BOSS) and FS+BAO (BOSS+small-z).
 }
 \end{table}

\paragraph{Patchy mocks} In Fig.~\ref{fig:patchy_wcdm} and Table~\ref{tab:patchy_wcdm}, we show the marginalized posterior distribution of the $w$CDM parameters with a BBN prior on $\omega_b$, obtained by fitting the mean over 2048 CMASS NGC Patchy mocks, analyzed with the covariance for one box divided by 16. 
We also show there the results obtained by adding to the fit `small-$z$' priors on $r_s/D_V$ at $z_{\rm eff}=0.106$ and on $D_V/r_s$ at $z_{\rm eff}=0.15$ similar to 6DF and SDSS DR7 MGS respectively, but with central value on the truth of the simulation and error bars rescaled by 4.

Let us discuss first the case without small-$z$.
By measuring the theory-systematic error as the distance to the truth of the $68\%$ confidence intervals of the 1D posteriors of the fit to the mean over the patchy mocks, we can detect a theory-systematic error within $\sim 1/3 \sigma_{\rm stat}$ of the BOSS data.
Although we detect no significant theory error with respect to the error bars obtained in the BOSS data, with at most a marginal one in $\ln ( 10^{10} A_s)$ of $\lesssim 1/3 \sigma_{\rm stat}$, we observe that the trends in the shifts of the posteriors between $\Lambda$CDM (+$\nu$) and $w$CDM are similar for both patchy and BOSS data, resulting in slightly lower $A_s$ and $n_s$, and slightly higher $\omega_{\rm cdm}$ and $H_0$.
This can be traced to the anti-correlations among $w$CDM in the FS+BAO analysis, as discussed in sec.~3.
In particular, we find that $w$ prefers, although not significantly, a lower value than $-1$, driving the trends we are observing. 

Adding the small-$z$ prior, we find that the posteriors are pushed towards the truth values, leading to even smaller theory errors, that become negligible. This is expected as adding another redshift helps breaking degeneracies.
 
\paragraph{Lettered challenge simulations} 

The BOSS `lettered' challenge boxes are N-body simulations of side length $2.5 \, {\rm Gpc}/h$ and are described in e.g.~\cite{DAmico:2019fhj}.
We consider two independent realizations of the lettered challenge simulations. 
The first realization consists in 4 boxes populated by 4 different halo occupation distribution (HOD) models, labelled A, B, F, and G, and the other one, labelled D, is populated by yet another HOD model. 
We measure the theory-systematic error using those simulations as follows.
We fit A, B, F, and G separately and average the posteriors of the cosmological parameters over the boxes, and fit D separately.
Using one of these two realizations, we measure the theory-systematic error on a cosmological parameter as the distance of the 68\%-confidence interval of the 1D posterior to the truth.
In particular, if the truth lies within the 68\%-confidence region, the theory-systematic error is zero.
Furthermore, as ABFG and D are independent realizations, we can combine them. 
The combination of ABFG+D allows us to measure the theory systematics using a volume about $14$ times larger than the BOSS effective volume.
In practice, we combine the 1D posteriors of the shifts of the mean from the truth (as the product of two Gaussians).
For each cosmological parameter, the theory-systematic error is then given as the distance of the 68\%-confidence interval of the resulting 1D posteriors of the shifts to zero. 
Notice that this represents a conservative requirement given the number of cosmological parameters we actually measure.
The precision of our measurements of the theory-systematic error using this technique is given by the error bars obtained by the combination of ABFG+D.

\begin{figure}[h]
\centering
\includegraphics[width=0.49\textwidth]{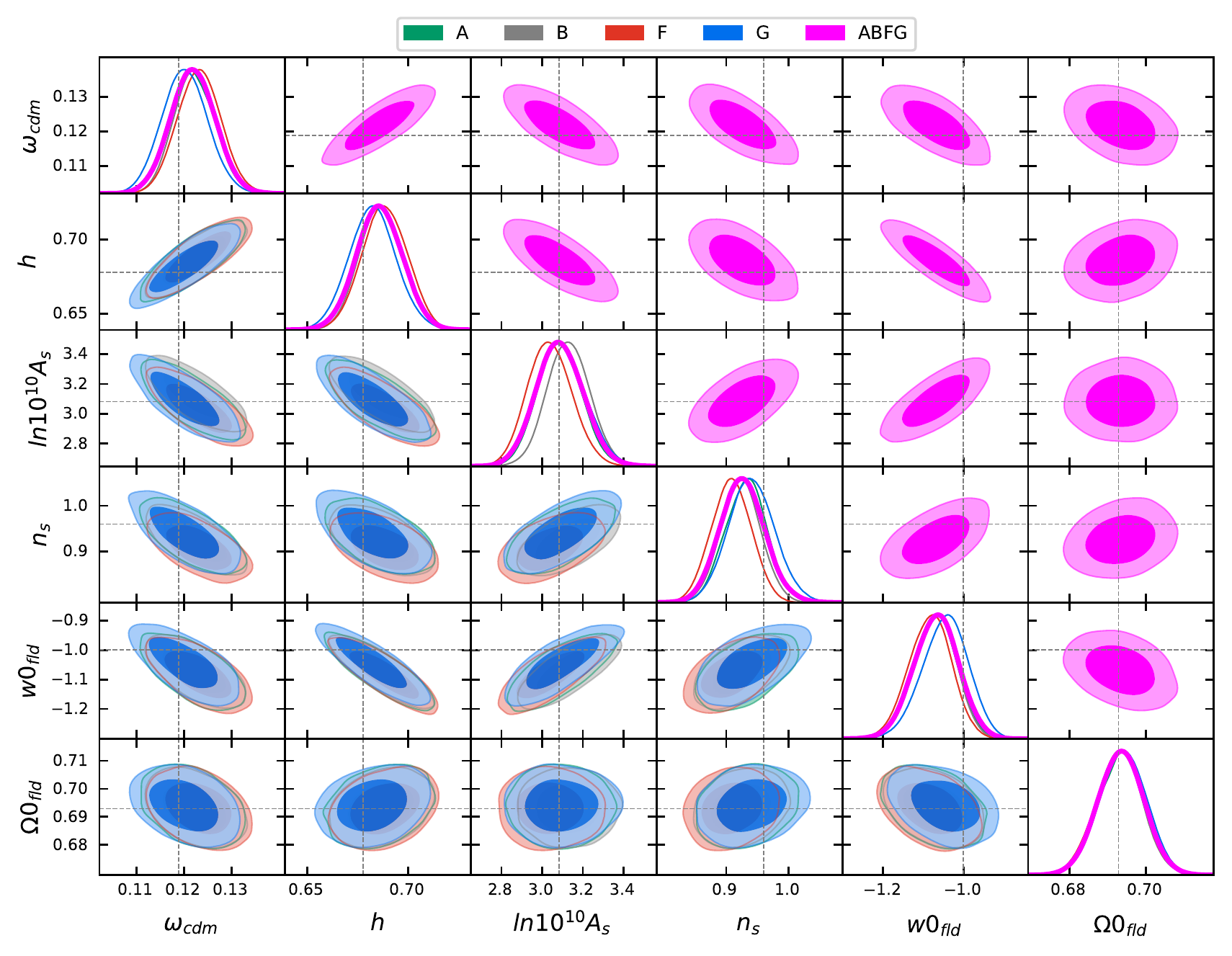}
\includegraphics[width=0.49\textwidth]{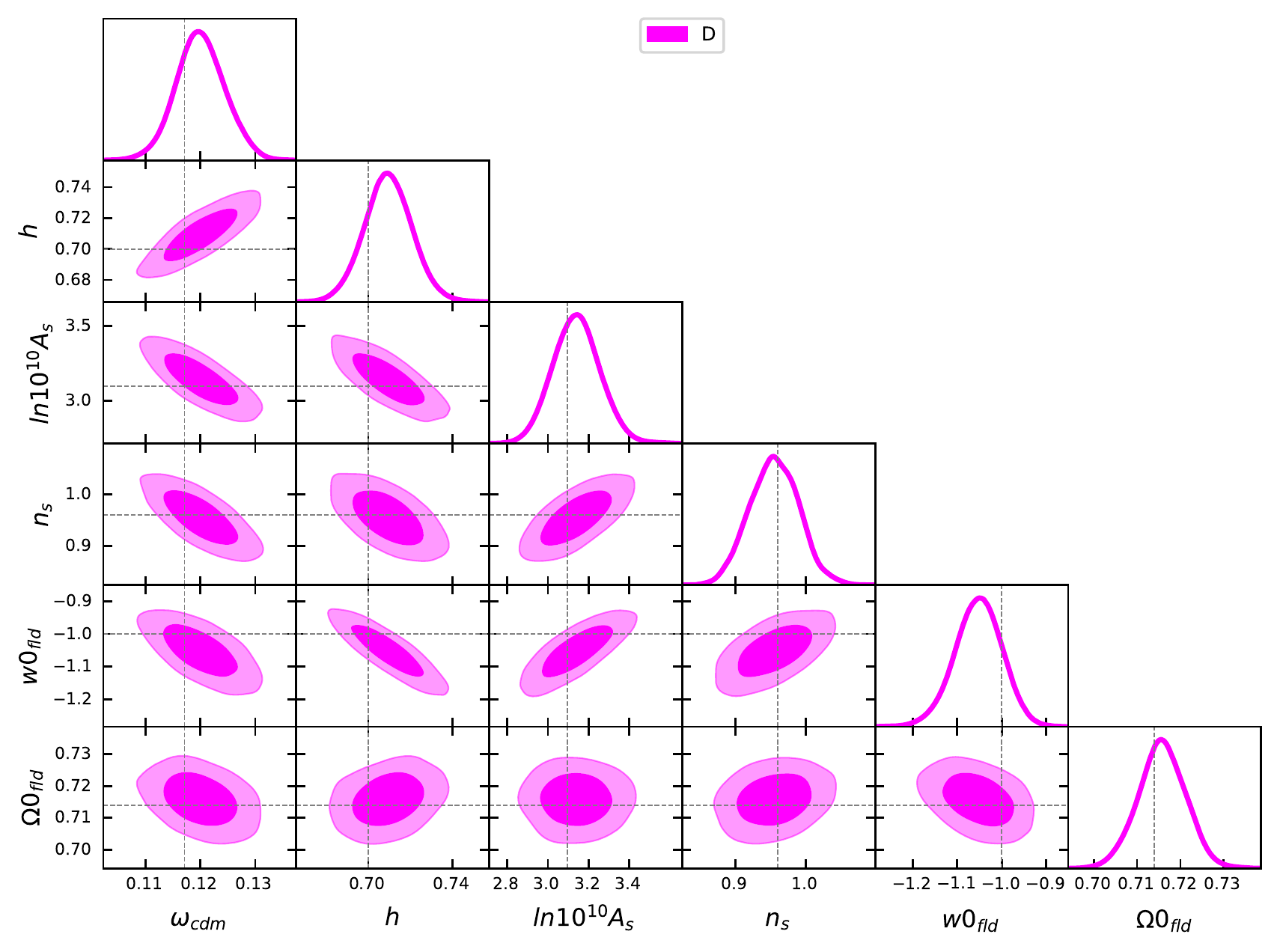}
\caption{\small Triangle plots obtained fitting the lettered challenge simulations on $w$CDM with a BBN prior and BAO priors.
The dashed lines represent the truth of the simulations.}
\label{fig:challenge}
\end{figure}

For the analyses we are concerned with in this work, the BOSS FS is always fit jointly with BAO, either from the same redshift bins (reconstructed BOSS BAO) or from other redshifts (small-$z$ and Lyman-$\alpha$ BAO), in order to get significant constraints on $w$ (see discussions in sec.~\ref{sec:physics}). 
We do not have post-reconstructed measurements from the lettered challenge simulations, and so we analyze them with the following `BAO' Gaussian priors, centered on the true cosmology of the simulations, and with width equal to the error bars of the respective experiments divided by 4~(\footnote{The factor of 4 is roughly the square root of the volume ratio of the challenge simulations with BOSS.}):
\begin{itemize}
\item a `small-$z$' prior on $r_s/D_V$ at $z_{\rm eff}=0.106$ and on $D_V/r_s$ at $z_{\rm eff}=0.15$, inspired by 6DF and SDSS DR7 MGS, with widths $0.00375$ and $0.04$, respectively ;
\item a `Lyman-$\alpha$' prior on $D_H/r_s$ and $D_A/r_s$ at $z_{\rm eff}=2.34$, inspired by eBOSS Lyman-$\alpha$ auto-correlation and cross-correlation to quasars, with widths $0.055$ and $0.3125$, respectively (see Eqs.~(50-51)~in~\cite{Blomqvist:2019rah}).
\end{itemize}
Notice that this procedure is allowed because these data sets are uncorrelated with the ones at BOSS redshift.
The 1D and 2D posteriors are shown in Fig.~\ref{fig:challenge}. 
The 68\%-confidence intervals and the theory-systematics are given in Table~\ref{tab:challenge}.
We detect zero theory-systematic errors except small ones on $h$ and $w$ of $0.001$ and $0.022$, respectively, which are, compared to the error bars obtained on the data in Table~\ref{tab:wcdm_lss}, less than $1/5 \, \sigma_{\rm data}$, thus negligible.
Similar conclusions hold for the case without the `Lyman-$\alpha$ prior, where we get only a negligible theory-systematic error on $w$ of $0.035$, which is less than $1/4 \, \sigma_{\rm data}$.
These tests on simulations allow us to confidently analyze $w$CDM on the observational data.

\begin{table}[h]
\footnotesize
\centering
\begin{tabular}{l|c|c|c|c|c|c} 
 \hline 
 		
  		& $\omega_{cdm}$ & $h$ & $\ln (10^{10} A_s)$ & $n_s$ & $w$ & $\Omega_m$				\\ 
& $\sigma_{\rm stat} | \sigma_{\rm sys}$ & $\sigma_{\rm stat} | \sigma_{\rm sys}$ & $\sigma_{\rm stat} | \sigma_{\rm sys}$ & $\sigma_{\rm stat} | \sigma_{\rm sys}$ & $\sigma_{\rm stat} | \sigma_{\rm sys}$ & $\sigma_{\rm stat} | \sigma_{\rm sys}$   \\ \hline   
ABFG 		& $0.005 | 0.000 $ & $0.011 | 0.000$ & $0.11 | 0.00$ & $0.035 | 0.000$ & 0.055 | 0.013 & 0.006 | 0.000	 \\ \hline 
D			& $0.005 | 0.000 $ & $0.011 | 0.000$ & $0.11 | 0.00$ & $0.035 | 0.000$ &	0.052 | 0.000 & 0.005 | 0.000	 \\ \hline 
ABFG+D		& $0.003 | 0.000 $ & $0.008 | 0.001$ & $0.08 | 0.00$ & $0.025 | 0.000$ &	0.038 | 0.022 & 0.004 | 0.000	 \\ \hline 
ABFG w/o Ly-$\alpha$		& $0.007 | 0.000 $ & $0.012 | 0.000$ & $0.14 | 0.00$ & $0.040 | 0.000$ & 0.057 | 0.037 & 0.009 | 0.000	 \\ \hline 
D w/o Ly-$\alpha$  			& $0.006 | 0.000 $ & $0.012 | 0.000$ & $0.12 | 0.00$ & $0.039 | 0.000$ &	0.059 | 0.005 & 0.008 | 0.000	 \\ \hline 
ABFG+D  w/o Ly-$\alpha$		& $0.005 | 0.000 $ & $0.009 | 0.000$ & $0.09 | 0.00$ & $0.028 | 0.000$ &	0.041 | 0.035 & 0.006 | 0.000	 \\ \hline 
 \end{tabular}
 \caption{\small 68\%-confidence intervals $\sigma_{\rm stat}$ and theory-systematic errors $\sigma_{\rm sys}$ obtained fitting the lettered challenge simulations on $w$CDM with a BBN prior and BAO priors. }
 \label{tab:challenge}
\end{table}

\section{Marginalized likelihood and best fit}
\label{app:marginalized}

In this appendix, we explicitly show that even using the likelihood analytically marginalized over some of the EFT parameters one can recover with good accuracy their best fit values, necessary to recover the best fit power spectrum.
We start by expressing the theory model as a sum of terms multiplied by EFT parameters appearing linearly plus all the other terms:
\begin{equation}
    P_{\alpha} = \sum_i b_{G,i} P^i_{G,\alpha} + P_{NG,\alpha} \, .
\end{equation}
Here we use a concise notation in which the index $\alpha$ runs over $k$-bins and multipoles; $b_{G,i}$ are the EFT parameters over which the marginalization is analytical, and both $P^i_{G,\al}$ and $P_{NG,\al}$ will depend on the cosmological parameters and the non-linear EFT parameters which cannot be analytically integrated out.
The posterior can then be written as
\begin{equation}
     -2 \ln \mathcal{P} = (P_{\al} - D_{\al} ) 
     C^{-1}_{\al \bt} ( P_{\bt} - D_{\bt} )
     + b_{G, i} \sigma^{-1}_{ij} b_{G, j} - 2 \ln \Pi \, ,
\end{equation}
where $D_\al$ is the data vector, $C_{\al \bt}$ is the data covariance, and we introduced a Gaussian prior on the $b_{G, i}$ with covariance $\sigma_{ij}$, plus a generic prior $\Pi$ on the cosmological and non-linear EFT parameters.

Collecting different powers of $b_{G,i}$, the posterior can be written in the form:
\begin{equation}
    \label{eq:nonmargP}
    -2 \ln \mathcal{P} = b_{G,i} F_{2, ij} b_{G,j} - 2 b_{G,i} F_{1, i} + F_0 \, ,
\end{equation}
where we defined the following terms:
\begin{align}
    & F_{2, ij} = P^i_{G, \al} C^{-1}_{\al \bt} P^j_{G, \bt} + \sigma^{-1}_{ij} \, , \\
    & F_{1, i} = - P^i_{G, \al} C^{-1}_{\al \bt} (P_{NG, \bt} - D_{\bt}) \, , \\
    & F_0 =  (P_{NG, \al} - D_{\al}) C^{-1}_{\al \bt} (P_{NG, \bt} - D_{\bt}) - 2 \ln \Pi \, .
\end{align}

Performing a Gaussian integral on the $b_{G, i}$, we obtain the marginalized posterior:
\begin{equation}
    \label{eq:margP}
    -2 \ln \mathcal{P}_{\rm marg} = - F_{1,i} F^{-1}_{2,ij} F_{1,j} + F_0 + \ln \det \left(\frac{F_2}{2 \pi}\right) \, .
\end{equation}
We will now show the relation between the extrema of eq.~\eqref{eq:nonmargP} and of eq.~\eqref{eq:margP}.
Setting the gradients of eq.~\eqref{eq:nonmargP} to zero we find:
\begin{align}\label{eq:firsteqbg}
    & b_{G,i} = F^{-1}_{2,ij} F_{1,j} \, , \\
    & b_{G,i} \frac{\de F_{2, ij}}{\de c_n} b_{G,j} - 2 b_{G,i} \frac{\de F_{1, i}}{\de c_n} + \frac{\de F_0}{\de c_n} = 0 \, ,
\end{align}
where $c_n$ is any other parameter we want to fit.
The first equation is already solved in terms of the $c_n$.
Substituting into the second, we find the nonlinear equation which determines the best fit for the $c_n$ using the non-marginalized likelihood:
\begin{equation}
\begin{split}
    0 &= F_{1,k} F^{-1}_{2,ik} \frac{\de F_{2, ij}}{\de c_n} F^{-1}_{2,jl} F_{1,l} - 2 \frac{\de F_{1, i}}{\de c_n} F^{-1}_{2,ij} F_{1,j} + \frac{\de F_0}{\de c_n} \\
    &=
    - F_{1,i} \frac{\de F^{-1}_{2, ij}}{\de c_n} F_{1,j} - 2 \frac{\de F_{1, i}}{\de c_n} F^{-1}_{2,ij} F_{1,j} + \frac{\de F_0}{\de c_n} \, ,
\end{split}
\end{equation}
where we used the fact that $F_{2,ij}$ is symmetric and we expressed its derivative in terms of the derivative of its inverse.

If we instead start from eq.~\eqref{eq:margP}, to find the best fit, we can use eq.~(\ref{eq:firsteqbg}) for the $b_{G,i}$'s best fit, while  the best fit of the $c_n$ parameters is given by the solution of the following equation:
\begin{equation}
    - F_{1,i} \frac{\de F^{-1}_{2, ij}}{\de c_n} F_{1,j} - 2 \frac{\de F_{1, i}}{\de c_n} F^{-1}_{2,ij} F_{1,j} + \frac{\de F_0}{\de c_n} + F^{-1}_{2, ji} \frac{\de F_{2, ji}}{\de c_n} = 0 \, .
\end{equation}
Because of the last term, the best fit point using the marginalized posterior is shifted with respect to the best fit point obtained from the non-marginalized posterior.

However, this term is generically small. To get an idea of the size of the various terms, let us consider a single $b_G$ and a single $k$-bin.
We can estimate the size of the last and the first terms as $F_2^{-1} \frac{\de F_2}{\de c_n} \sim P_{G}^{-1} \frac{\de P_G}{\de c_n}$, $F_1 \frac{\de F^{-1}_2}{\de c_n} F_1 \sim \frac{(P_{NG}-D)^2}{P_G} C^{-1} \frac{\de P_G}{\de c_n}$.
Therefore
\begin{equation}
\left| \frac{F^{-1}_{2, ji} \frac{\de F_{2, ji}}{\de c_n}}{F_{1,i} \frac{\de F^{-1}_{2, ij}}{\de c_n} F_{1,j}} \right| \sim \frac{C}{(P_{NG}-D)^2} \,.
\end{equation}
Now, $(P_{NG}-D)^2$ is of the size of the one-loop squared, while $C$ is the squared error on the data, which at $k_{\rm max}$ is of order of the two-loop squared, therefore the r.h.s. is negligible.

In practice, we find that the best fits from the marginalized posterior and from the non-marginalized posterior are equal to better than $0.1\%$ precision on all parameters for all analyses we performed on data or simulations.

\bibliographystyle{JHEP}
\bibliography{references}

\end{document}